\documentclass[a4paper,fleqn]{cas-dc}

\usepackage[square,numbers]{natbib}

\usepackage{todonotes}

\usepackage{caption}
\usepackage{subcaption}

\usepackage{url}

\usepackage{tikz}
\newcommand*\emptycirc[1][1ex]{\tikz\draw (0,0) circle (#1);} 
\newcommand*\halfcirc[1][1ex]{%
  \begin{tikzpicture}
  \draw[fill] (0,0)-- (90:#1) arc (90:270:#1) -- cycle ;
  \draw (0,0) circle (#1);
  \end{tikzpicture}}
\newcommand*\fullcirc[1][1ex]{\tikz\fill (0,0) circle (#1);}

\usepackage{tabularx}
\usepackage{multirow}

\newcolumntype{K}{>{\raggedleft\arraybackslash}X}
\newcolumntype{B}{K}
\newcolumntype{M}{>{\hsize=.75\hsize}K}
\newcolumntype{S}{>{\hsize=.5\hsize}K}
\newcolumntype{T}{>{\hsize=.25\hsize}K}

\newcolumntype{E}{X}
\newcolumntype{N}{>{\hsize=.75\hsize}X}
\newcolumntype{U}{>{\hsize=.5\hsize}X}
\newcolumntype{V}{>{\hsize=.25\hsize}X}

\newcolumntype{Y}[1]{>{\RaggedLeft\arraybackslash}p{#1}}

\newcommand{\heading}[1]{\multicolumn{1}{c}{#1}}

\newcommand\fix[1]{{#1}}
\newcommand\typo[1]{{#1}}

\newcommand\textas[1]{\texttt{#1}}

\usepackage{adjustbox}
\usepackage{tabto}

\makeatletter
\def\paragraph{\@startsection{paragraph}{4}{1\parindent}{0pt plus 3pt}%
{-1\parindent}{\normalfont\normalsize\bfseries}}%
\makeatother

\usepackage{floatrow}

\def\tsc#1{\csdef{#1}{\textsc{\lowercase{#1}}\xspace}}
\tsc{WGM}
\tsc{QE}
\tsc{EP}
\tsc{PMS}
\tsc{BEC}
\tsc{DE}

\begin{document}
\let\WriteBookmarks\relax
\def\floatpagepagefraction{1}
\def\textpagefraction{.001}

\shorttitle{Tor AS-level Adversaries}

\shortauthors{Gegenhuber et~al.}

\title [mode = title]{An Extended View on Measuring Tor AS-level Adversaries}

\author[1]{Gabriel K. Gegenhuber}

\ead{gabriel.gegenhuber@univie.ac.at}

\credit{Conceptualization, Methodology, Software, Writing - Review \& Editing}

\affiliation[1]{organization={University of Vienna, Research Group Security and Privacy},
    addressline={Kolingasse 14-16}, 
    city={Vienna},
    postcode={1090}, 
    country={Austria}}

\author[1]{Markus Maier}
\ead{markus.maier@univie.ac.at}
\credit{Software, Validation, Data Curation, Visualization}

\author[1]{Florian Holzbauer}
\ead{florian.holzbauer@univie.ac.at}

\credit{Data Curation, Validation, Writing - Original Draft}

\affiliation[2]{organization={SBA Research},
    addressline={Floragasse 7}, 
    city={Vienna},
    postcode={1040}, 
    country={Austria}
}

\affiliation[3]{organization={Christian Doppler Laboratory for Security and Quality Improvement in the Production System Lifecycle, University of Vienna},
    addressline={Kolingasse 14-16}, 
    city={Vienna},
    postcode={1090}, 
    country={Austria}
}

\author%
[2]
{Wilfried Mayer}
\ead{wmayer@sba-research.org}
\credit{Conceptualization, Methodology, Software}

\author%
[2]
{Georg Merzdovnik}
\ead{gmerzdovnik@sba-research.org}
\credit{Supervision}

\author%
[1]
{Edgar Weippl}
\ead{edgar.weippl@univie.ac.at}
\credit{Supervision}

\author%
[3]
{Johanna Ullrich}
\ead{johanna.ullrich@univie.ac.at}
\credit{Writing - Review \& Editing, Supervision, Project administration, Funding acquisition}

\begin{abstract}
Tor provides anonymity to millions of users around the globe which has made it a valuable target for malicious actors.
As a low-latency anonymity system, it is vulnerable to traffic correlation attacks from strong passive adversaries such as large autonomous systems (ASes).
In preliminary work~\cite{mayer2020actively},
we have developed a measurement approach utilizing the RIPE Atlas framework --
a network of more than 11,000 probes worldwide --
to infer the risk of deanonymization for IPv4 clients in Germany and the \typo{US}.

In this paper, we apply our methodology to additional scenarios providing a broader picture of the potential for deanonymization in the Tor network.
In particular,
we (a) repeat our \fix{earlier (2020)} measurements \fix{in 2022} to observe changes over time,
(b) adopt our approach for IPv6 to analyze the risk of deanonymization when using this next-generation Internet protocol,
and (c) investigate the current situation in Russia, where censorship has been intensified after the beginning of Russia's full-scale invasion of Ukraine.
According to our results, Tor provides user anonymity at consistent quality:
While individual numbers vary in dependence of client and destination,
we were able to identify ASes with the potential to conduct deanonymization attacks.
For clients in Germany and the \typo{US},
the overall picture, however, has not changed since 2020.
In addition, the protocols (IPv4 vs. IPv6) do not significantly impact the risk of deanonymization.
Russian users are able to securely evade censorship using Tor.
Their general risk of deanonymization is, in fact, lower than in the other investigated countries.
Beyond, the few ASes with the potential to successfully perform deanonymization are operated by Western companies,
further reducing the risk for Russian users.

\end{abstract}

\begin{highlights}
\item Update on Tor AS-level adversaries (comparison of measurement results for IPv4 in 2020 and 2022)
\item Measuring AS-level adversaries for IPv6
\item Measuring AS-level adversaries in censored countries (case study, based on popular clients and blocked destinations in Russia)
\end{highlights}

\begin{keywords}
Tor \sep RIPE Atlas \sep Traceroute measurements \sep Censorship \sep Privacy \sep Anonymity \sep Routing
\end{keywords}

\maketitle

\section{Introduction}\label{sec:introduction}
Tor is the most notable anonymity network, used by two to three million people every day.
A total of 6,500 voluntarily operated Tor relays advertise up \fix{to} 700 Gbit/s of bandwidth,
and provide anonymity by rerouting traffic via three Tor nodes.
As a low\typo{-}latency network, Tor is prone to traffic correlation attacks;
thereby, a malicious actor must be able to observe the traffic between the client originating the \typo{connection} and the first Tor node as well as the traffic between Tor's exit node and the destination.
A global passive observer is capable to do so,
but this form of \typo{an} attacker is explicitly excluded from Tor's threat model.
Yet, powerful observers exist, potentially threatening the anonymity of Tor users.
Their capabilities are, however, not exactly clear.
One reason for this is the theoretical assumption that the underlying Internet hierarchy is flat and evenly distributed. 
This is not the case, as the Internet is shaped in different tiers as well as various entities with different levels of control, e.g., Internet Exchange Points (IXP) with a high level of control and smaller Internet Service Providers (ISPs) with a lower level of control.
Also, the Tor network does not utilize the Internet in an evenly distributed manner as the location of Tor relays is depending on various external parameters, e.g., economical (the price of bandwidth) or political (censorship, prosecution) reasons.

Prior work~\cite{feamster2004location, edman2009awareness, nithyanand2016measuring} has shown that Tor traffic takes only a limited set of routes on the Internet.
These studies, however, rely on BGP updates and route prediction,
and claim that measurements -- despite being more reliable -- would be infeasible due to lacking measurement nodes in the autonomous systems (ASes) that host Tor users, nodes, and destinations. 
With the introduction of the RIPE Atlas framework~\cite{staff2015ripe} --
a global measurement network with more than 11,000 probes --
this assumption no longer holds. 
In our preliminary work~\cite{mayer2020actively},
we developed a measurement methodology utilizing this network to actively probe the Tor network.
In more detail, 
we used the probes to traceroute the Internet paths that are taken by Tor traffic
and, based on the collected data, estimated the correlation potential of AS-level adversaries.
\fix{In comparison to BGP-based approaches of path prediction, active measurements based on tracerouting reveal how the packets are actually routed over the Internet. This provides a more realistic risk estimation for Tor users as BGP-based approaches are known to overestimate their risk~\cite{juen2015defending}.}

The paper at hand is an extended version of our preliminary work~\cite{mayer2020actively}:
We apply our methodology to three additional use cases creating an extended view on AS-level adversaries.
In particular,
we (a) repeat our measurements from 2020 to observe changes in Tor's service quality over time,
(b) adopt our approach for IPv6 to analyze the threat of deanonymization when using this next-generation Internet protocol,
and (c) investigate the current situation in Russia as censorship has been intensified since the beginning of its full-scale invasion of Ukraine, starting on February 24th, 2022.
More specifically, the contributions of this paper are as follows:
\begin{description}
  \item \textbf{Updated View on AS Interconnections.} 
  By repeating our measurements from 2020, we investigate whether economical or political factors impacted Tor's service quality.
  Like in our previous measurements, 
  we identified a few ASes with the potential to successfully deanonymize Tor users;
  although individual numbers vary over time,
  the overall picture has remained unchanged.
  According to our results, Tor provides anonymity at a constant quality to its users in Germany and the \typo{US}.

 \item \textbf{AS-level Adversaries in IPv6.} 
  We are the very first to conduct active measurements investigating the status quo of IPv6 in the Tor network.
  Despite the fact that the number of IPv6 Tor relays is smaller than their IPv4 counterparts,
  we could not identify an increased threat of deanonymization for clients using Tor over IPv6,
  neither in Germany or the \typo{US}, nor in Russia.
  
  \item \textbf{Censorship in Russia.}
  With Russia's full-scale invasion in Ukraine, Russian state authorities also intensified Internet censorship, i.e., blocking media outlets reporting on the ongoing war.
  We investigated whether Russian clients evading censorship with Tor are prone to deanonymization, particularly when accessing blocked destinations in their geographic proximity.

\end{description}

The remainder of the paper is organized as follows:
Section~\ref{sec:background} provides background on Tor,
and Section~\ref{sec:relatedwork} discusses related work.
Section~\ref{sec:methodology} explains our measurement methodology.
Section~\ref{sec:evaluation} provides our measurements' results which are then discussed in 
Section~\ref{sec:discussion}.
We \fix{outline the limitations of our work in Section~\ref{sec:limitations} and}
draw our final conclusions in Section~\ref{sec:conclusion}.

\section{Background}
\label{sec:background}
Tor was designed by Dingledine~\cite{dingledine_tor_2004} in 2004
and soon became the most popular anonymity system. 
Tor's protocol specifications are open source and updated on a regular basis~\cite{tor2022spec}.

\begin{figure}[t]
	\centering
	\includegraphics[width=.95\textwidth]{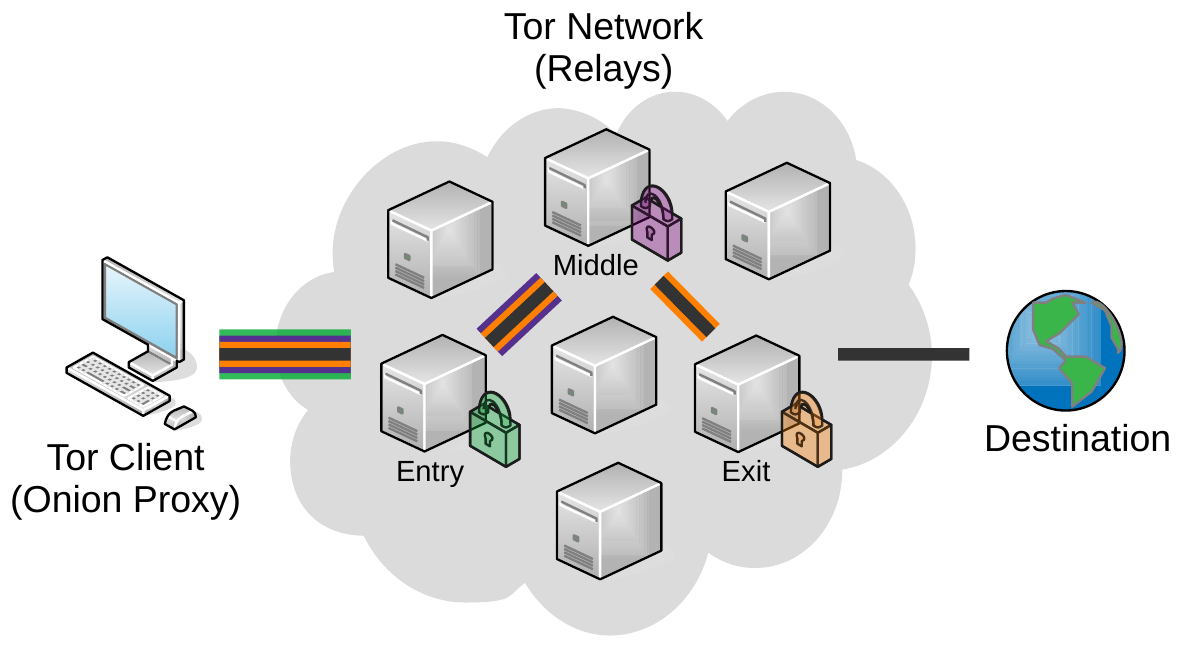}
	\caption{Tor network. Traffic is relayed via three Tor nodes to hinder correlation of the client and the destination. }
	\label{fig:tor}
\end{figure}

\textbf{Functionality.}
Tor is a low-latency anonymity network based on onion routing. 
\typo{It} forms an overlay network of at least three relay nodes that are used to detour user traffic.
The entrance to the Tor network is established by the onion proxy, also referred to as the \emph{Tor client}. The proxy handles connections from user applications and is responsible for fetching the initial network information about the Tor network from a set of trusted directory servers. 
This information is then used to select Tor nodes for relaying. 
The first relay along a Tor path is called the \emph{guard relay} -- it is the only one that \typo{knows} the client's IP address.
The last one on the path is the \emph{exit relay} which is the only one that \typo{knows} the target IP address.
The design of the Tor network is shown in Figure~\ref{fig:tor}.

\textbf{Path Selection.}
For path selection, the onion proxy relies on information retrieved from the directory servers. 
The information includes relay flags and bandwidth information about Tor nodes. 
The exit node is selected first,
then the guard relay, and finally the middle relay. 
The guard- and exit relays are selected randomly;
however, the relays are weighted by their bandwidth.
The middle relay is selected from the remaining set of nodes.
To protect the users and maximize their anonymity, guard- and exit relays are reused according to a strict ruleset (e.g., guard pinning).
Additionally, directory servers ensure that only nodes fulfilling certain uptime- and bandwidth requirements are selected as guard nodes.  
Another requirement for path selection is that the nodes have to belong to different /16 IPv4 prefixes. 
In reaction to new threat models, these rules are updated frequently, see also Section~\ref{sec:relatedwork}.

\textbf{Deanonymization of Users.}
Tor's design makes it vulnerable to a global passive observer,
which monitors all traffic going to and coming from the anonymity network.
Such a global observer is explicitly excluded from Tor's threat model;
however, powerful observers exist and threaten user anonymity.
If an entity is able to monitor both the incoming and outgoing packets of a communication channel,
it is able to correlate traffic entering Tor with traffic exiting the network based on timing.
Our work precisely focuses on this threat
and estimates probabilities of individual ASes
appearing in a client's entry- and exit path.

\textbf{IPv6 Support.}
Currently, Tor relays are either operated IPv4 only %
or Dualstack (i.e., providing an IPv4 and an IPv6 address).
This way,
Tor allows IPv6 traffic to enter and exit the network.
Thereby, Tor relays can also act as bridges between IPv4 and IPv6.
It should be noted that connecting from or to an IPv6 address reduces the set of possible relay candidates on the respective connection endpoint.

\section{Related Work}\label{sec:relatedwork}

Feamster and Dingledine~\cite{feamster2004location} provided the first analysis of location diversity in the Tor network for independently operated ASes based on BGP routing tables.
They analyzed the probability of an entry path to the network and an exit path from the network crossing through the same AS.
Their analysis showed that previous methods of choosing paths/nodes based on IP prefixes are not sufficient to guarantee a diverse set of ASes, since there was a 10\% to 30\% chance, that both the entry and exit path to
the mix network crossed the same AS. 
A refinement of this approach by Edman and Syverson in 2009~\cite{edman2009awareness} showed that the previous study had even underestimated the potential threat.
A study of Tor security properties against traffic correlation attacks was presented by Johnson et al.~\cite{johnson2013users}.
Their results showed that, depending on location, a user's chance of compromise can be at 95\% within three months of monitoring against a single AS.
One mitigation they proposed is to carefully select which entry and exit nodes to use.
Wacek et al.~\cite{wacek2013empirical} built a graph of the Tor network to capture the networks' AS boundaries. 
Using this graph they provided an evaluation of a set of proposed relay selection methods and quantified their respective anonymity properties. 
Their results showed that bandwidth is an important property for the performance of such algorithms, and should not be neglected.

The importance of location diversity in the Tor network has been shown by several attacks proposed in recent years.
Vanbever et al.~\cite{vanbever2014anonymity} provided a study of the capabilities of AS-level adversaries.
Sun et al.~\cite{sun2015raptor} described a set of advanced routing attacks on Tor, named \textit{Raptor}.
They also described the feasibility of asymmetric AS-level attacks by observing not only data traffic from the exit relay to the server but also TCP acknowledgment traffic on other routes which increases the capabilities of AS-level adversaries.
Including the reverse path, they found 31.7\% of the Tor circuits to be vulnerable in their measurements.
However, paths had different probabilities to be selected by a client,
and the actual number was likely to be lower.
In 2016, Nithyanand et al.~\cite{nithyanand2016measuring} also used data on the Internet's topology~\cite{giotsas2014inferring} in a combination with AS-topology simulations~\cite{gill2012modeling} to estimate the threat posed by adversaries to Tor users. 
While previous attempts at the correlation of traffic~\cite{hopper2010much, mittal2011stealthy} had very limited performance or required a large amount of captured traffic or time, \textit{DeepCorr}~\cite{nasr2018deepcorr}, developed by Nasr et al. greatly improved the feasibility of such attacks.
By leveraging emerging learning mechanisms they managed to achieve drastically higher performance compared to existing state-of-the-art systems.

To mitigate the threat of \typo{AS-level adversaries that are able to correlate traffic and thereby} monitor Tor users, various kinds of protection mechanisms have been proposed~\cite{alsabah2016performance}.
Nithyanand et al. proposed \emph{Astoria}~\cite{nithyanand2016measuring}, an AS-aware Tor client.
While similar in functionality to \textit{LASTor}~\cite{akhoondi2012lastor}, it provided improved protection with concern to threat models and attacker capabilities.
Sun et al.~\cite{sun2017counter} presented a measurement study on the security of Tor against BGP hijacking attacks and presented a new relay selection mechanism to mitigate such attacks on Tor.
In contrast to previous approaches, \emph{DeNASA} from Barton et al.~\cite{barton2016denasa} provided a mechanism for AS-aware path selection independently of the destination.
Additionally, they proposed another system for the creation of efficient and anonymous Tor circuits~\cite{barton2018towards}.
Hanley et al.~\cite{hanley2019dpselect} proposed an extension to the work presented by Sun et al.~\cite{sun2017counter} to increase the provided privacy and anonymity guarantees.
Wan et al.~\cite{wan2019guard} showed that several attacks against a set of the proposed protections (Counter-RAPTOR, DeNASA, and LASTor) were still possible, but they also proposed simple solutions, which allowed to mitigate the threat posed by their developed methods.
Rochet et al.~\cite{rochet_claps_2020} introduced client-location-aware path selection (CLAPS) to overcome the pitfalls detected in previous path selection solutions (Counter-Raptor, DeNASA). They proved that based on the path selection of the earlier approaches the client's location can be revealed only after a few connections. Eaton et al.~\cite{ateniese_improving_2022} further enhanced the receiver side anonymity of Tor by introducing Private Information Retrieval (PIR) to hide which information is retrieved from the Hidden Service directory servers (HSDirs).
Next to security, recent related work also focused on improving the Tor core. Jansen et al.~\cite{hohlfeld_accuracy_2021} estimated that the actual bandwidth of the Tor network could be much higher. They suggested a new measurement system for bandwidth calculation of Tor nodes. The authors found that with the current system the bandwidth self-measurements resulting in the \emph{observed bandwidth} are rather imprecise.

\begin{figure*}[ht]
	\centering
	\includegraphics[width=.95\textwidth]{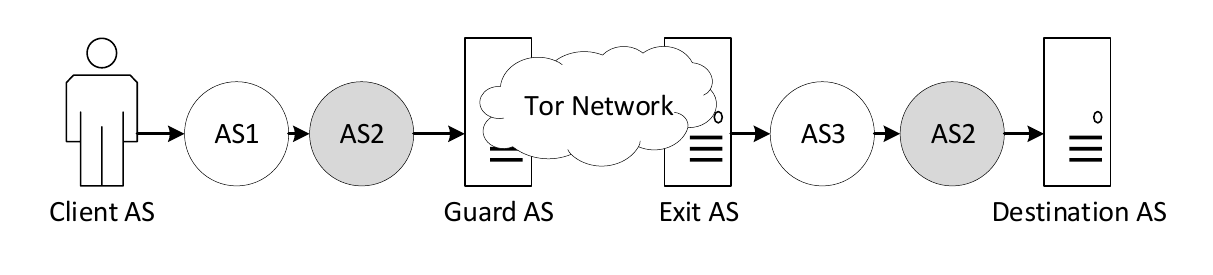}
	\caption{Threat model. AS2 appears on the Tor entry path, between the client and the guard relay,
	and on the exit path, between the exit relay and the destination, and is thus in a position to perform traffic correlation deanonymizing the client.}
	\label{fig:correlation-attack}
\end{figure*}

\begin{table*}[t]
    \begin{subtable}{\textwidth}
    \begin{subtable}[h]{0.38\textwidth}
        \def\arraystretch{1}
        \begin{tabularx}{0.98\columnwidth}{V S S S}
            \toprule
                 & 2020   & 2022 & Diff\\ 
            \midrule%
            All   & 6,509    &   6,559  & \textcolor{ForestGreen}{+1\%} \\ %
            Exit  & 1,000    &   1,597  & \textcolor{ForestGreen}{+60\%} \\  %
            Guard & 2,415    &   2,272  & \textcolor{BrickRed}{-6\%} \\ %
            \bottomrule
        \end{tabularx}
        \caption{Relays}
        \label{tab:tor.relays}
    \end{subtable}
     \hfill
    \begin{subtable}[h]{0.30\textwidth}
        \def\arraystretch{1}
        \begin{tabularx}{0.98\columnwidth}{S S S}
            \toprule
                2020   & 2022 & Diff \\ 
            \midrule%
            1104 &  981 & \textcolor{BrickRed}{-11\%} \\    %
            275  &  222 & \textcolor{BrickRed}{-19\%} \\  %
            470  &  469 & \textcolor{BrickRed}{-0\%} \\  %
            \bottomrule
        \end{tabularx}
        \caption{Diff. ASes}
        \label{tab:tor.as}
    \end{subtable}
    \begin{subtable}[h]{0.3\textwidth}
        \def\arraystretch{1}
        \begin{tabularx}{0.98\columnwidth}{S S S }
            \toprule
            2020   & 2022 & Diff\\ 
            \midrule%
            418  & 694 & \textcolor{ForestGreen}{+66\%} \\
            113  & 181 & \textcolor{ForestGreen}{+60\%} \\
            255  & 368 & \textcolor{ForestGreen}{+44\%} \\
            \bottomrule
        \end{tabularx}
        \caption{Bandwidth (GBit/s)}
        \label{tab:tor.bw}
    \end{subtable}
    
    \end{subtable}
    
    \caption{Tor relay statistics. While the number of relays increased, they are now spread among fewer ASes. The total Tor bandwidth increased by 66\% over the past two years.}
    \label{tab:tor}
\end{table*}

\section{Methodology}\label{sec:methodology}
In the following section, we describe our method to measure strong AS-level observers, which are in a good position to conduct correlation attacks.
As an overlay network, Tor depends on the underlying structure of the Internet. 
While often a flat hierarchy is assumed, it is clear that this is not the case.
We can model the structure of the Internet by looking at autonomous systems identified by a unique AS number (ASN).
One AS can be seen as an administrative entity that is responsible for a defined routing policy. 
Some AS are large and include a lot of Tor users, destinations, or relays, others do not contain users and destinations but are used for routing Tor traffic through the Internet and others are not important for Tor routing at all.
Thus, some entities can observe more traffic than others.

With our measurement approach, we find a way to quantify which entities are in a stronger position. 
Figure~\ref{fig:correlation-attack} illustrates the basic idea of a standard traffic correlation attack, where one adversary (AS2) is placed on the incoming route to Tor as well as on the outgoing route to the destination.
Sun et al.~\cite{sun2015raptor} showed that it is also possible to correlate reverse-path traffic that may be routed differently.
Other work already quantified strong adversaries with the help of BGP route updates. 
In contrast, we develop a method that utilizes the RIPE Atlas framework to actively acquire routing information \fix{as this allows to study how packets actually travel the Internet}.

The paper at hand extends our preliminary work;
therefore, we apply our measurement approach to three additional use case\typo{s} to gain a broader view o\typo{f} the potential of deanonymization in the Tor network.
(a) We repeat the measurements and compare the state of 2020 with the current state (September 2022). 
(b) As IPv6 support at Tor relays has improved over the recent years~\cite{tor2021ipv6}, 
we adapt our methodology to additionally acquire routing information for IPv6.
(c) Lastly, we investigate a practical case study, Russia's full\typo{-}scale invasion \typo{of} Ukraine,
and analyze AS-level packet routing by simulating access to websites that are blocked by Russian state authorities.

\subsection{Relay AS Diversity}\label{sec:relay-diversity}
As shown in Table~\ref{tab:tor}, the Tor network currently (September 17th, 2022) consists of 6,559 \fix{relays}.
Only relays with the \emph{Guard} flag (stable and reliable relays after a ramp-up phase~\cite{torblog2013lifecycle}), are used as entry relays.
Only relays configured to allow exiting traffic are potential exit relays in a Tor circuit.
Because of the more stringent requirements, the number of guard and exit relays (with guard/exit probability $>0$) is smaller than 6,559.
This also affects the AS diversity, which is the number of different ASes these relays are placed in.

Table~\ref{tab:tor} compares the metrics of Tor relay nodes for our two measurement snapshots in 2020 and 2022.
Overall, the network has grown in terms of size and offered bandwidth.
However, it has become more centralized, as for example, the AS diversity at exit relays has decreased by nearly 20\%.
Although the number of guard relays dropped by 6\%, the number of bridges -- providing an alternative and more anonymous entry to the network -- nearly doubled (1,350 vs. 2,450) in the respective time period.
Since a Tor relay node can -- additionally to its IPv4 address -- also offer an IPv6 address, we give an overview of the current IPv6 support in Table~\ref{tab:tor-ipv6}.
With IPv6, the AS diversity drops by more than 60\% making it an interesting target for our study.

\begin{table*}[t]
 \begin{subtable}{\textwidth}
       \begin{subtable}[h]{0.39\textwidth}
        \centering
        \def\arraystretch{1}
        \begin{tabularx}{0.98\columnwidth}{V M M S}
            \toprule
                 & All   & IPv6 & Share\\ 
            \midrule%
            All & 6,559 & 2,924 & 45\% \\ %
             Exit & 1,597 & 1,083 & 68\% \\  %
            Guard & 2,272 & 951  & 42\% \\ %
            \bottomrule
        \end{tabularx}
        \caption{Relays}
        \label{tab:tor-ipv6.relays}
    \end{subtable}
     \hfill
    \begin{subtable}[h]{0.30\textwidth}
        \centering
        \def\arraystretch{1}
        \begin{tabularx}{0.98\columnwidth}{S S S}
            \toprule
            All   & IPv6 & Share\\ 
            \midrule%
            981 & 375 & 38\% \\    %
            222 & 94& 42\% \\  %
            469 & 175 & 37\%  \\  %
            \bottomrule
        \end{tabularx}
        \caption{Diff. ASes}
        \label{tab:tor-ipv6.as}
    \end{subtable}
    \begin{subtable}[h]{0.3\textwidth}
        \centering
          \def\arraystretch{1}
        \begin{tabularx}{0.98\columnwidth}{S S S }
            \toprule
            All   & IPv6 & Share \\ 
            \midrule%
            694 & 342 & 49\% \\
            181 & 128 & 71\% \\
            368 & 152 & 41\% \\
            \bottomrule
        \end{tabularx}
        \caption{Bandwidth (GBit/s)}
        \label{tab:tor-ipv6.bw}
    \end{subtable}
    
    \end{subtable}
    
    \caption{IPv6 support statistics. As of Sept. 2022, 45\% of the Tor relays support IPv6, while the exit bandwi\typo{d}th is 71\% of the IPv4's one.}
    \label{tab:tor-ipv6}
\end{table*}

\begin{figure*}
	\centering
    \includegraphics[width=\textwidth]{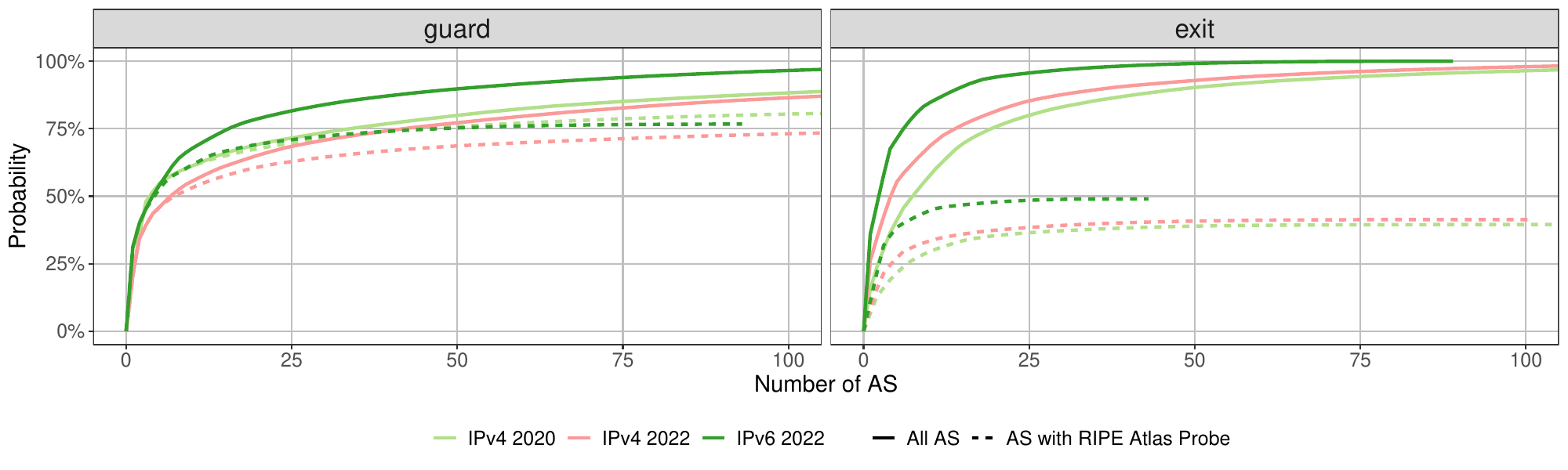}
	\caption{Accumulated percentage of (a) guard, and (b) exit probability with the number of ASes. While RIPE probes cover 75\% of guard probability, they cover less than 50\% of exit probability. }
	\label{fig:as-guard-propability}
\end{figure*}

Tor relays are chosen based on their flags and consensus weight.
In Figure~\ref{fig:as-guard-propability}, we show the AS diversity relation to guard and exit probability. 
We see that a small number of \typo{ASes have} a large share of (a) guard and (b) exit probability. 
For IPv4, only five ASes control more than 50\% of exit probability and 43 ASes have more than 90\%.
We also see that six ASes have \typo{ a summarized guard probability of more than 50\%} and 131 have more than 90\%.
During our measurements in 2020, half of the exit probability was controlled by eight ASes and only four ASes dominated half of the guard probability.
Therefore, the accumulated exit probabilities among top ASes \fix{has become} even more centralized, while the guard probabilities are now slightly more diversified.
\fix{For IPv6 the centralization is even worse, as
only three ASes control more than 50\% of guard resp. exit probability.
}
Summarizing, Tor relays are distributed in almost 1000 ASes, 
the majority of entry and exit routing endpoints are however placed in a few ASes only.

Location diversity provides a similar picture:
Two countries (Germany and \typo{the US}) account for more than 47\% of the relays.
While the top five countries are still the same as in 2020,
we noticed that Russia has lost a majority of its relays and has dropped from the sixth to the 18th rank (from 297 down to just 65 relays).

\begin{table}
    \begin{tabularx}{\textwidth}{S X X K K K K}
    \toprule
    & \heading{AS}    & \heading{Name} & \heading{Relays} & \heading{Gbit/s} & \heading{$P_{exit}$} & \heading{$P_{guard}$} \\ 
    \midrule
   \multirow{5}{*}{\rotatebox{90}{Exit}} & \small{60729} & \tiny{ZWIEBELFR.}       &        225 &   39.2 &      .221 &       .002 \\
    & \small{205100} & \tiny{F3NETZE}       &         32 &   11.9 &      .084 &       .000 \\
    & \small{200651} & \tiny{FlokiNET}      &         48 &   5.95 &      .030 &       .000 \\
    & \small{62744} & \tiny{QUINTEX}       &        100 &   6.77 &      .026 &       .000 \\
    &  \small{4224} & \tiny{CALYX}         &         29 &   5.36 &      .023 &      .001 \\
    \hline
   \multirow{5}{*}{\rotatebox{90}{Guard}} & \small{201814} & \tiny{SKYTECH}       &         81 &  13.62 &      .023 &       .018 \\
    & \small{46844} & \tiny{SHARKTECH}     &         36 &   7.17 &      .000 &       .014 \\
    & \small{19437} & \tiny{SS-ASH}        &         10 &   2.51 &      .000 &       .006 \\
    & \small{200303} & \tiny{LUMASERV}      &         20 &   2.54 &      .000 &       .006 \\
    & \small{264617} & \tiny{PANAGLOBAL}    &          5 &   1.83 &      .000 &       .005 \\
    \bottomrule
    \end{tabularx}
	\caption{AS hosting Tor relays but no RIPE Atlas probe. Adding a single probe to \textas{AS60729} would increase the accumulated exit probability by 22.1\%. }
	\label{tab:as-without-ripe}
\end{table}
\subsection{The RIPE Atlas Framework}\label{sec:ripe}

\begin{figure}[t]
	\centering
	\includegraphics[width=.95\textwidth]{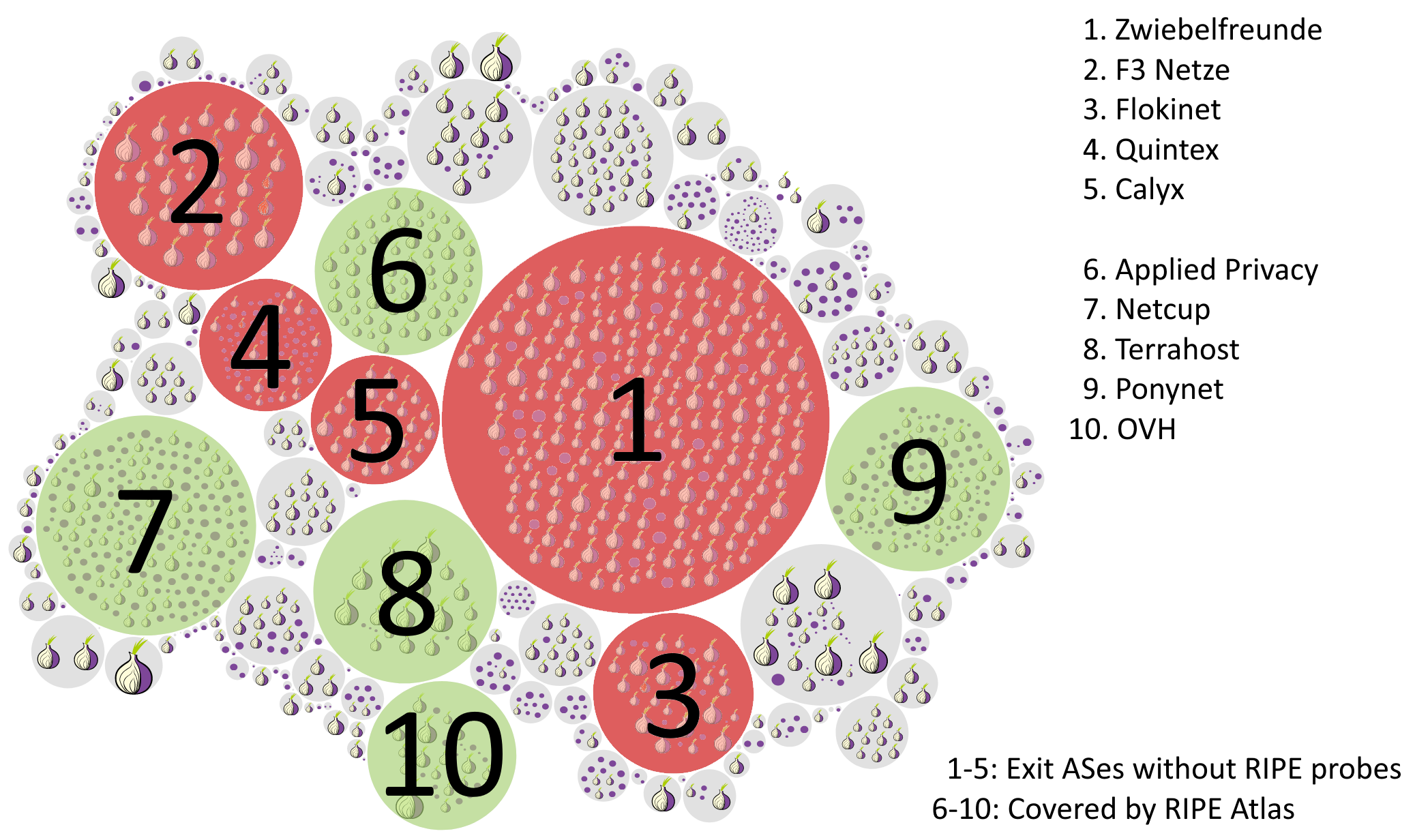}
	\caption{RIPE Atlas probe coverage of ten large exit relay ASes. Adding a few probes to the exit relay ASes in \fix{dark red} color \fix{(numbers 1 - 5)} could significantly increase the coverage \cite{torbubbles2022}}
	\label{fig:exit-relays-without-ripe}
\end{figure}
The RIPE Atlas framework is a highly distributed measurement network consisting of more than 11,000 available probes, deployed in over 3,600 different ASes.
Regarding IPv6, it offers more than 5,000 vantage points (i.e., probes) in over 1,600 different ASes.

The measurement platform allows us to execute various low-level commands, e.g., ping or traceroute, on these probes and further processes the results. 
We will utilize this to execute traceroute commands from RIPE Atlas probes that are deployed in the same ASes as Tor guard- or exit relays, as well as clients and popular destinations.

Figure~\ref{fig:as-guard-propability} also shows the cumulated guard- and exit probability for ASes that contain RIPE Atlas probes. 
From 222 ASes that contain exit relays, only 98 also contain a probe (837 relays out of 1,597).
Still, that makes approx. 43\% of the total exit probability (35\% with only 12 ASes).
This differs from the cumulated guard probability. 
From 469 ASes that contain 2,272 relays, 249 ASes (with 1,723 relays) also include a RIPE Atlas probe, which represents guard relays with a sum of 80\% guard probability (60\% with 15 ASes).
Especially for exit relays, these numbers could be drastically increased if only a few, exit-focused ASes would also host RIPE Atlas probes.
Table~\ref{tab:as-without-ripe} identifies ASes, that are currently not hosting any RIPE probes.
By adding only five probes we could measure ASes with 81\% exit probability in total and ten probes would reach 88\% probability in total.

Figure~\ref{fig:exit-relays-without-ripe} shows a bubble graph of current exit relays sorted by AS.
The top five ASes that are not covered by RIPE Atlas (cf. Table~\ref{tab:as-without-ripe}) are marked in red (numbers 1-5).
\textas{AS208323 APPLIEDPRIVACY} which is represented by a green bubble (6) was missing RIPE Atlas coverage in 2020, but now hosts a RIPE probe.
Other exemplary exit ASes that are covered by RIPE Atlas are also marked in green (numbers 7 - 10).
Compared to 2020, the overall RIPE Atlas coverage of ASes that contain guard- and exit relays \fix{has not changed much} (exit: 41 to 43\%; guard: 83 to 80\%).

\subsection{Active \emph{traceroute} Probing with RIPE Atlas}\label{sec:probing}
\begin{figure*}
	\centering
	\includegraphics[width=.87\textwidth]{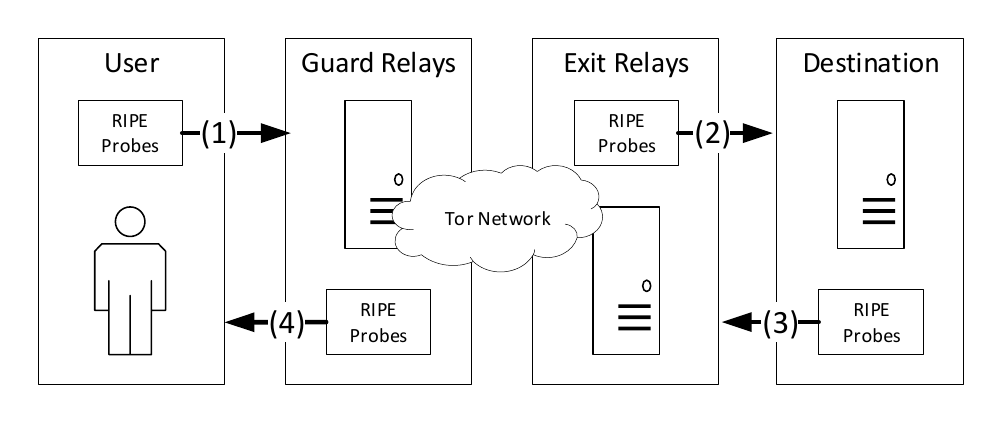}
    \vspace{-3mm}
	\caption{RIPE Atlas \emph{traceroute} scans. The forward path is covered by $D_1$, from client to guard relay ASes, and $D_2$, from exit relay ASes with probes to the destination AS. The reverse path is covered by $D_3$, from destinations to exit relay ASes with ripe probes, and $D_4$, from guard relay ASes with probes back to the client.}
	\label{fig:scans}
\end{figure*}

As illustrated in Figure~\ref{fig:scans} we perform \emph{traceroute} measurements to identify routes taken for four different directions: 
(1) all client ASes to all guard ASes, 
(2) exit ASes with probes installed to the destination ASes, 
(3) destination ASes to all exit ASes, and 
(4) guard ASes with probes installed to the client ASes. 
With these measurements, we do not cover all possible routes since not all ASes have probes installed. 
For IPv4, depending on the direction, we measure \fix{at step} (1) 100\%, (2) $\sim$43\%, (3) 100\%, and (4) $\sim$80\% in terms of route probability.
For IPv6, we cover \fix{at step} (1) 100\%, (2) $\sim$52\%, (3) 100\%, and (4) $\sim$85\% in terms of route probability.

In detail, this process works as follows:

\begin{enumerate}
    \item Create the following sets:
    \begin{enumerate}
        \item[i.] $AS_{client}$ \tabto*{2.2cm} ... ASes of the clients (as chosen for the individual scenario, see Subsection~\ref{sec:origin})
        \item[ii.] $AS_{guard}$ \tabto*{2.2cm} ... all ASes with guard relays %
        \item[iii.] $AS_{guard \cap probe}$ \tabto*{2.2cm} ... all ASes with guard relays and RIPE Atlas probes %
        \item[iv.] $AS_{exit}$ \tabto*{2.2cm} ... all ASes with exit relays %
        \item[v.] $AS_{exit \cap probe}$ \tabto*{2.2cm} ... all ASes with exit relays and RIPE Atlas probes %
        \item[vi.] $AS_{destination}$ \tabto*{2.2cm} ... ASes of the destinations (as chosen for the individual scenario, see Subsection~\ref{sec:origin}) \vspace{1mm}
    \end{enumerate}
    \item Generate ICMP traceroute measurement definitions for the following directions: \vspace{1mm}
    
    \begin{enumerate} 
        \item[(1)] \parbox{2.2cm}{$AS_{client}$}       $\xrightarrow{traceroute} \enspace AS_{guard}$ 
        \item[(2)] \parbox{2.2cm}{$AS_{exit \cap probe}$}   $\xrightarrow{traceroute} \enspace AS_{destination}$
        \item[(3)] \parbox{2.2cm}{$AS_{destination}$}  $\xrightarrow{traceroute} \enspace AS_{exit}$  
        \item[(4)] \parbox{2.2cm}{$AS_{guard \cap probe}$}  $\xrightarrow{traceroute} \enspace AS_{client}$
    \end{enumerate} \vspace{2mm}
    \item Execute the \emph{traceroute} with the RIPE Atlas measurement API (\texttt{"protocol": "ICMP", "response\_timeout": 20000, "packets": 1}).

    \item Process all results and look up the corresponding AS from the \emph{ip2asn} database.
    \item For every \emph{traceroute}, mark all included ASes with the probability of that path being chosen, i.e., the corresponding guard/ exit probability.
    \item Combine the values for \typo{directions} 1 and 4 for the entry side, and 2 and 3 for the exit side, s.t., if an AS appears on either the forward or the reverse path it is assigned with the probability of that path being chosen. For multiple destinations, all traceroutes are combined.
    \item Point out the top ASes, that appear on entry and exit side by looking at $P_{guard} \cap P_{exit}$.
\end{enumerate}

\subsection{Origin and Destination AS}\label{sec:origin}
The goal of this work is to investigate multiple Tor scenarios for their proneness to \fix{deanonymization} attacks. Therefore, we (a) repeat the measurements of our preliminary work in 2020 to observe changes over time,
(b) adopt our approach for IPv6 to analyze the threat when using Tor over the next-generation Internet protocol,
and (c) extend our measurements to investigate the current situation in Russia as censorship has intensified after its full-scale invasion of Ukraine, starting on February 24th, 2022.
The following paragraphs describe how we derived the ASes for our client and destination data sets.

\textbf{IPv4 Measurements.}
As mentioned in Section~\ref{sec:relay-diversity} Germany and the \typo{US} account for nearly half of all Tor nodes.
In our 2020 measurements, we have therefore chosen the ten ASes in Germany and the \typo{US} containing most RIPE Atlas probes -- an indicator of the AS's popularity in the respective country -- for the client set $C$.
For destinations, we derive the ASes from the Tranco~\cite{LePochat2019Tranco} top sites list.
In particular, we take the 100 top-ranked domains, resolve the domain, and retrieve the corresponding ASes.
We include only those ASes with deployed RIPE Atlas probe(s) in our destination set $D$.
For our \emph{traceroute} measurements, we select one RIPE Atlas probe for each AS in the client and destination set.

For the repetition of our measurements in 2022, we slightly adapted the approach in the following manner:
In addition to the ASes inferred according to the just described procedure, 
we also included ASes that have been investigated in the first iteration (i.e., $C_{ipv4} = C_{2022} \cup C_{2020}$ and $D_{ipv4} = D_{2022} \cup D_{2020}$).
For some cases, we were not able to gather updated results for ASes that have been measured in 2020.
For example, \textas{AS3356 LEVEL 3} was included in our destination set in 2020, but was not measured in 2022 as it does \fix{not} host a RIPE Atlas probe anymore.
\fix{
Similarly, the 2020 client set contains historic ASes that do not exist nowadays (e.g., two consumer-grade ASes \textas{AS6830} and \textas{AS31334} were merged to \textas{AS3209} -- which is already present in our client set).
}

\textbf{IPv6 Measurements.}
For clients, we again select ten ASes in Germany and \typo{the US} with the most RIPE Atlas probes offering IPv6 support.
For destinations, we increased the number of included domains from the Tranco list from 100 to 250 due to overall low IPv6 support.
For comparison, we also included all ASes supporting IPv6 in the IPv4 datasets and vice versa.

\textbf{Websites Blocked by Russia.}
Russian ISPs had started to block Tor in December 2021~\cite{russia2021tor}, i.e., three months before the beginning of Russia's full-scale invasion of Ukraine on February 24th 2022.
Afterwards, Russia introduced even more rigorous censoring, blocking access to social media and independent news outlets~\cite{russia2022russia}.
Many of the blocked destinations are hosted either in Russia or Ukraine and report on the \typo{ongoing} war.
Circumvention of censorship is among the main goals of Tor,
making Russia's full-scale invasion of Ukraine an interesting case study.
Thus, we investigate whether users from Russian client ASes could be deanonymized when accessing these censored destinations in their geographical proximity.

For the client set, we again determine the ASes with the most RIPE Atlas probes in the respective country, i.e., Russia.
For the destination set, we use a public list of websites blocked by Russian state authorities~\cite{russia2022ukraine} and rank the domains by popularity using the Tranco list. Then, we resolve these domain names and filter for ASes in Russia or Ukraine.
Finally, we match our results with the RIPE Atlas deployment which determines the destination set for this measurement.
As there were only two AS candidates supporting IPv6 within this data set, we refrained from a distinct IPv6 measurement in this particular case.

\textbf{Summary of Data Set.}
In total, we have eight data sets representing client ASes:
2020 IPv4 Germany, 2020 IPv4 \typo{US}, 2022 IPv4 Germany, 2022 IPv4 \typo{US}, 2022 IPv4 Russia, 2022 IPv6 Germany, 2022 IPv6 \typo{US}, 2022 IPv6 Russia.
Please note that there is no 2020 data set for Russia as the country was not included in our previous measurements.
Beyond, we have four data sets representing destination ASes: 2020 IPv4 Tranco, 2022 IPv4 Tranco, 2022 IPv6 Tranco, 2022 IPv4 Blocked Websites.
A detailed list of the ASes that are included in the data set is found in the Appendix~\ref{sec:appendix-client-and-destination-sets}.

Our approach allows to measure Tor's entry and exit paths independently of each other, and to combine the results only in a successive processing step. 
Thus, it is sufficient to measure each of the data sets only once.

\subsection{Data Sources}\label{sec:data}
To facilitate reproducibility and encourage openness, all used data files are publicly available at the project 
website\footnote{Project website: \url{https://github.com/sbaresearch/ripe-tor}}.
In particular, our work relies on following data sources:

\begin{enumerate}
    \item The Tor consensus, that contains all Tor relays with their IP addresses (IPv4, IPv6), associated flags (particularly "Guard" and "Exit"), advertised bandwidth, and guard and exit probability.
      We collect this information via the Tor network status protocol \emph{onionoo}\footnote{onionoo: \url{https://metrics.torproject.org/onionoo.html}}.
    \item Statistical data about the \emph{RIPE Atlas probes}\footnote{probes: \url{https://atlas.ripe.net/probes/}}. 
      We use different data (e.g., id, number and AS of the probes) to find all probes connected to the same ASes as guard and exit relays.
    \item Freely accessible \emph{ip2asn}\footnote{ip2asn: \url{https://iptoasn.com/}} databases to match IP addresses with the corresponding AS number.
    \item Active RIPE Atlas \emph{traceroute} results\footnote{measurements: \url{https://atlas.ripe.net/measurements}}. 
   
\end{enumerate}

\section{Evaluation}\label{sec:evaluation}
In this section, we discuss the results of our measurements:
For readability, we first illustrate our approach in an exemplary measurement including a single client and destination AS only (see Subsection~\ref{sec:singlescan}).
Then, we focus on the full measurement discussing the ASes residing on Tor's entry paths (see Subsection~\ref{sec:evaluation_second}),
and those on the exit paths (see Subsection~\ref{sec:evaluation_third}).
Finally, we combine these results to infer ASes appearing on Tor's entry and exit path with high probability as they have the potential to perform traffic correlation deanonymizing Tor users (see Subsection~\ref{sec:evaluation_fourth}).

\subsection{Exemplary Measurement: Single Client and Destination}\label{sec:singlescan}
As an illustration of the capabilities of our methodology, we evaluate the results of measurements with a single fixed client AS and a fixed destination AS. 
Therefore, we choose the AS of our research center $C = \{AS1764\}$ as client, and the AS of one mirror of the \url{torproject.org} website $D = \{AS24940\}$ as destination.
In these ASes, we selected RIPE Atlas probes and scheduled 1,194 traceroutes, as defined in Section~\ref{sec:probing}; out of which 1,177 (98.6\%) were executed successfully (D1: 563/563, D2: 104/109, D3: 240/240, D4: 270/282).

Table~\ref{tab:single-result} shows the results for IPv4;
the ASes are grouped depending on whether they reside on a path towards a guard relay,
or on a path from an exit relay.
As expected, the client resp. destination AS (\textas{HETZNER}, \textas{NEXTLAYER}) is found in all traceroutes.
Beyond, the ASes \textas{HETZNER}, \textas{OVH} and \textas{ZWIEBELFREUNDE} appear in the tables;
serving a high share of guard resp. exit bandwidth in the Tor network,
the respective route is likely to be chosen as a Tor path ($P_{relays}$).
However, we want to focus on intermediate ASes,
that are different from those hosting relays as well as client/destination and appear on many routes.
We identified \textas{LEVEL3}, \textas{COGENT}, and \textas{HURRICANE} to be in a powerful position.

\begin{table}
\begin{tabularx}{\textwidth}{T N X S M M M M }
\toprule
 & \heading{AS}        & \heading{Name}          & C  & \heading{$P$}      & \heading{$P_{relays}$}     & \heading{$P_{routes}$}     & \heading{R} \\
\midrule
\multirow{5}{*}{\rotatebox{90}{entry}}
& {\small 1764 }  & {\tiny NEXTLAYER}  & {\tiny \emptycirc}   & 1.00  & -         & 1.00  & 468  \\ 
 & {\small 3356 }  & {\tiny LEVEL3}     & {\tiny \fullcirc}   & .330  & .002  & .328  & 59   \\ 
 & {\small 24940 } & {\tiny HETZNER}    & {\tiny \emptycirc}  & .224  & .224  & .000  & 2    \\ 
 & {\small 16276 } & {\tiny OVH}        & {\tiny \halfcirc}   & .131  & .130  & .000  & 2    \\ 
 & {\small 174 }   & {\tiny COGENT} & {\tiny \fullcirc}   & .123  & .002  & .121  & 110  \\ 
\midrule
\multirow{5}{*}{\rotatebox{90}{exit}} & 24940   & {\tiny HETZNER}  & {\tiny \emptycirc}    & 1.00  & -  & 1.00  & 220  \\ 
 & {\small 60729  } & {\tiny \fix{ZWIEBELFR.}}  & {\tiny \halfcirc}    & .221  & .221  & -         & 1    \\ 
 & {\small 25291 }  & {\tiny \fix{INTERDOTL.}}  & {\tiny \fullcirc}    & .221  & -         & .221  & 1    \\ 
 & {\small 47147  } & {\tiny AS-ANX}  & {\tiny \fullcirc}   & .162  & -         & .162  & 2    \\ 
 & {\small 6939 }   & {\tiny HURRICANE}   & {\tiny \fullcirc}   & .130  & .002  & .128  & 28   \\ 
\bottomrule
\end{tabularx}
\rule{0in}{1em}
{\tiny \emptycirc} Client or Destination AS.\hspace{1ex}
{\tiny \halfcirc} Guard or Exit AS.\hspace{1ex}\\
{\tiny \fullcirc} Transit AS.
\caption{Entry path and exit path probabilities for a single client and a single destination. \textas{HETZNER} appears on the entry path due to its high number of entry relays as well as on the exit path due to hosting the destination.}
\label{tab:single-result}
\end{table}

\begin{table}
\centering
\begin{tabularx}{\textwidth}{T T N X T S S S S}
\toprule
 & & \heading{AS} & \heading{Name} & C & \heading{$P_{guard}$} & \heading{$P_{exit}$}  & \heading{$P_{\&}$}  \\
\midrule
\multirow{3}{*}{\rotatebox{90}{2020}} & \multirow{3}{*}{v4} &  \small{24940}   & {\tiny HETZNER} & {\tiny \emptycirc}    & .202  & .988  & .199  \\ 
& & \small{1200}   & {\tiny AMS-IX1} & {\tiny \fullcirc}    & .180  & .068  & .012  \\ 
& & \small{16276}   & {\tiny OVH} & {\tiny \fullcirc}    & .152  & .065  & .010  \\ 
\midrule
 \multirow{5}{*}{\rotatebox{90}{2022}} & v4 &  \small{24940}   & {\tiny HETZNER} & {\tiny \emptycirc}    & .224  & 1.00  & .224  \\ 
\cline{2-8}
 & \multirow{4}{*}{v6}&  \small{24940}  & {\tiny HETZNER} & {\tiny \emptycirc}    & .350  & .998  & .350  \\
 & & \small{6939}    & {\tiny HURRICANE}  & {\tiny \fullcirc}   & .087  & .393  & .034  \\
 & & \small{47147}   & {\tiny AS-ANX ANE} & {\tiny \fullcirc}    & .107  & .223  & .024  \\
 & & \small{197540}  & {\tiny NETCUP-AS} & {\tiny \fullcirc}    & .107  & .139  & .015  \\
\bottomrule
\end{tabularx}
\rule{0in}{1em}
{\tiny \emptycirc} Client or Destination AS.\hspace{1ex}
{\tiny \fullcirc} Transit AS.
\caption{AS with the potential for traffic correlation. For all measurements, \textas{HETZNER} has the potential to deanonymize the client due to appearing on the entry and exit path.}
\label{tab:single-combined}
\end{table}
Eventually, we filter for intermediate ASes that have a probability of 1\% or higher to appear on both sides,
and only a single AS remains, namely \textas{HETZNER}.
With $P_{guard}=0.224$ and $P_{exit}=1.000$,
it has a chance of $P=0.224$ to \fix{deanonymize} Tor traffic from our research center to \url{torproject.org}.

Table~\ref{tab:single-combined} provides an overview of the ASes with a probability of 1\% or higher to appear on both sides for the three measurements 2020 IPv4, 2022 IPv4, and 2022 IPv6.
A comparison of the IPv4 measurements reveals that the number of such ASes decreased;
however, the probability of \textas{HETZNER} increased by 2.5 percentage points.
This means that the AS has now an even higher chance of \fix{deanonymization} due to its increased guard probability. 
For IPv6-based traffic, this number is even higher.
Because the set of possible guard relays decreases in IPv6, the guard probability of \textas{HETZNER} increases once again \fix{by more than 10 percentage points}.
Beyond, \typo{there} are three transit ASes in IPv6 with a $P_\&$ of up to 3.4\%.

The case of \textas{HETZNER} is particularly interesting as its chance of deanonymization arises from a distinct combination:
On the one hand, it is the destination of our measurements;
on the other hand, it hosts a high share of guard bandwidth and is thus more likely to be pinned as a guard node.
This raises the question of whether path selection should include the destination AS to prevent such scenarios.

\begin{figure*}
	\centering
    \includegraphics[width=\textwidth]{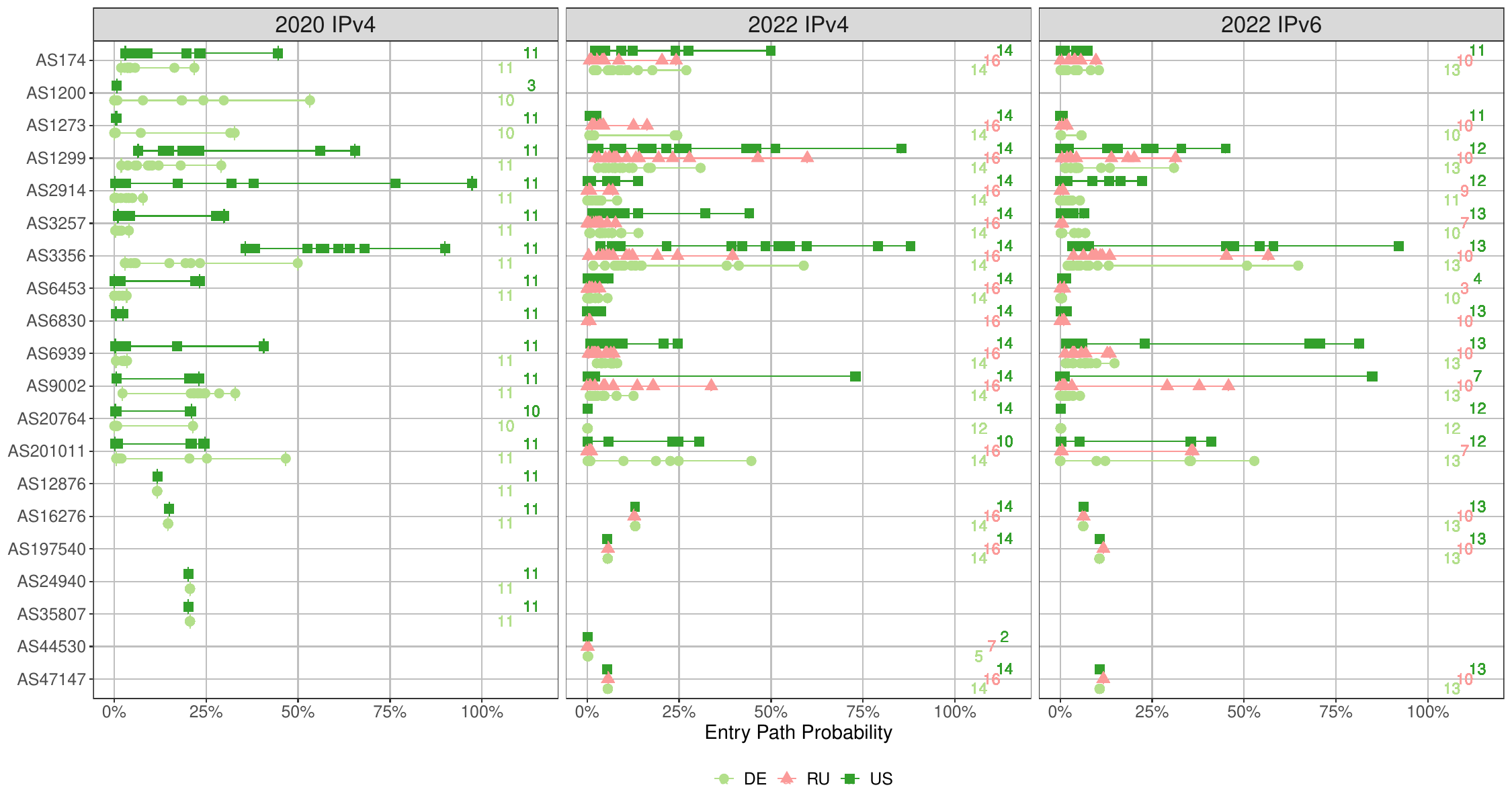}
  \caption{Entry path probability describing the chance of an AS to appear between the client AS of three different countries and the guard relay. Each data point represents a client AS. The number on the right is the total number of data points. }
	\label{fig:client-top-2022-total}
\end{figure*}

\subsection{Tor Entry: ASes between Clients and Guard Relays}
\label{sec:evaluation_second}
In the following paragraphs, we investigate the chance of ASes to be on a route to/from a guard relay and the chosen client ASes in Germany, \typo{the US}, and Russia.
For a total of 20 intermediate ASes, 
Figure~\ref{fig:client-top-2022-total} shows their entry path probabilities as inferred from our measurements.
The 20 AS were chosen according to the following rules:
For every country, we select the 15 most likely intermediary ASes.
We then show all intermediary ASes that occur for more than five clients in every measured country in our graph.
For graphs that correspond to measurements in 2022, we also include ASes that were selected at the previous measurement period.
Each data point represents one specific client AS. For a (transit) AS that is present in our measured routes to Tor guard relays, a data point in the graph denotes the summarized probabilities of all routes, i.e.,
the probability that this AS can trace packets from the client to the Tor network.
On the right side of each row, we show the total number of data points.
To visualize the range of the single data points, we draw a line between the minimum and maximum values.
The figure allows a comparison among our three measurements (2020 IPv4, 2022 IPv4, and 2022 IPv6) as well as among the chosen countries.

In essence, the overall picture for IPv4 was confirmed,
and the ASes with high entry path probability in 2022 remain the same as in 2020.
For example, any client AS uses -- though with varying probabilities -- paths including \textas{AS174 COGENT}, \textas{AS1299 TWELVE99}, \textas{AS3356 LEVEL3}.
Yet, certain changes were observed:
First,  \textas{AS1200 AMS-IX1}, \textas{AS12876 ONLINE S.A.S.}, and \textas{AS35807 SKYNET-SPB-AS} appeared in the 2020 measurements, but not in the latest of 2022. In return, previously unknown ASes were seen (\textas{AS44530 HOPUS HOPUS}, \textas{AS47147 AS-ANX ANEXIA}).
Second, \textas{AS2914 TT-COMMUNICATIONS-2} was frequently observed for US-based client ASes \fix{in 2020}; nowadays, its probability is roughly comparable to the Germany-based client ASes. 

\newlength\myHeight %
\newlength\myWidth
\begin{figure*}
	\centering

    \adjustbox{gstore width=\myWidth, gstore height=\myHeight, center}{
        \includegraphics[width=\textwidth]{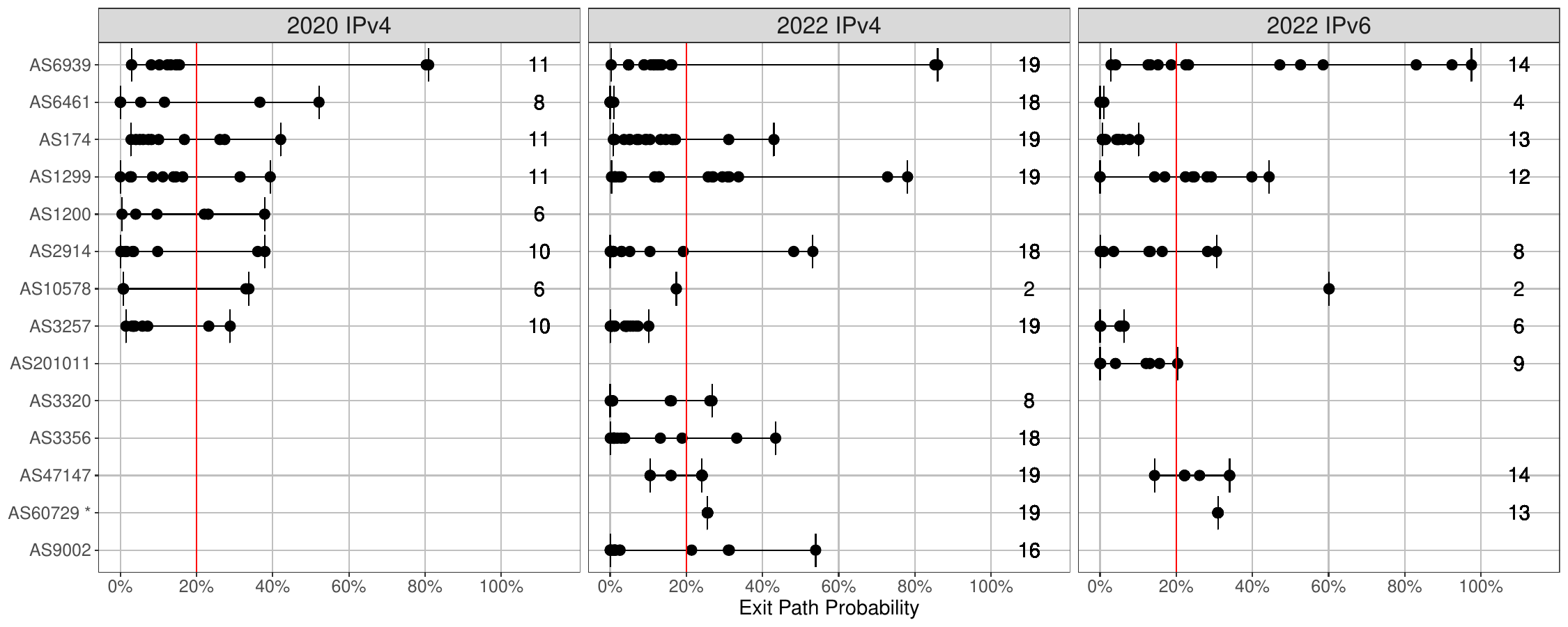}
    }
  \caption[Short caption for List of Figures without footnote]{Exit path probability describing the chance of an AS to appear between the exit relay and the Tranco list destinations. Each data point represents a destination AS. The number on the right is the total number of data points.}
	\label{fig:destination-top-2022-total}
\end{figure*}

Comparing IPv4 and IPv6, \textas{AS6939 HURRICANE} is found more often on paths from US-based client ASes;
in return, \textas{AS174 COGENT}, \textas{AS1299 TWELVE99} and \textas{AS3257 GTT-BACKBONE} are traversed less often.
Beyond this fact, 
Tor paths of IPv4 and IPv6 appear to be highly similar,
particularly for client ASes in Germany and Russia.

\textbf{Local Differences}
Most clients taking routes through high probability transit ASes are from the US:
For example, \textas{AS6939 HURRICANE} is particularly dominant for IPv6 in the \typo{US},
but plays only a minor role for German and Russian ASes.
\textas{AS9002 RETN-NET} plays a strong role in the \typo{US} but has lower probability in Russia \fix{and} Germany.
This might indicate that routing in the \typo{US} is more centralized than in the other countries.
Beyond, it appears that ASes that are frequently found on paths from German client ASes,
are also often seen on paths from Russian ASes;
this might be a consequence of their geographic proximity.

While the sole presence of an AS on a path to/from a guard relay is not sufficient to conduct traffic correlation,
it might however be sufficient to identify clients -- and successively \fix{their users} -- connecting to the Tor network.
Thus, the discussed results, covering Tor's entry side, also provide insights on which ASes are capable to detect clients using Tor.

\subsection{Tor Exit: ASes between Exit Relays and Destinations}
\label{sec:evaluation_third}
In the following paragraphs, we investigate the chance of ASes to be on a route between an exit relay and the chosen destinations for two distinct destinations sets: the Tranco List representing the most popular domains and those officially blocked by the Russian state-authority Roskomnadzor.

\textbf{Tranco List.}
For a total of 14 intermediary ASes,
Figure~\ref{fig:destination-top-2022-total} shows the probability for the destination ASes that have been inferred from the Tranco list.
For every destination we select all ASes that have a maximum probability of more than 20\%.
To filter for significant ASes we remove rows with a median value of less than 1\% or less than five data points.
For graphs that correspond to measurements in 2022, we also include ASes that were selected \typo{during} the previous measurement period.
Each data point represents a specific destination AS,
and the figure allows a comparison among our three measurements (2020 IPv4, 2022 IPv4, 2022 IPv6).
ASes that were selected because they are hosting a substantial amount of exit relays (e.g., \textas{AS60729 ZWIEBELFREUNDE}) are marked with an asterisk.

For the AS that have already been seen in the 2020 measurements,
we see a similar picture in 2022,
and only minor changes are apparent:
\textas{AS6461 ZAYO} is barely seen anymore,
and \textas{AS1200 AMS-IX1} is gone.
The latter has also been identified for the entry side.
Beyond, we found five new ASes with a considerable chance \fix{of being} along the path.
Comparing IPv4 and IPv6, we see that the maximum probabilities are typically lower for IPv6 for most ASes.
Conversely, \textas{AS6939 HURRICANE} has better chances to be on the path towards the destination,
i.e., this AS appears to be a dominant player in the IPv6 Internet.

\begin{figure} 
\centering
\includegraphics[width=.77\textwidth]{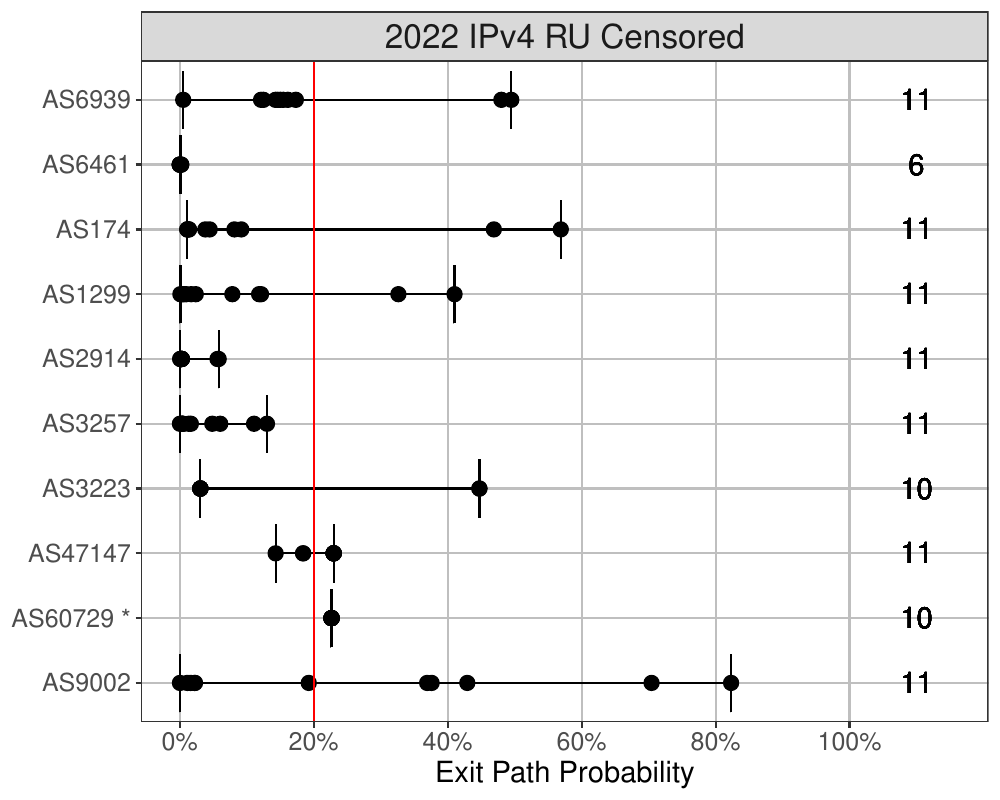}
\caption{Exit path probability for the ASes hosting websites that are blocked by Russia.
The depicted ASes are all operated by Western companies (i.e., US, SE, UK, AT, DE).
}
\label{fig:russia-censored-destination}
\end{figure}

\textbf{Destinations Blocked by Russia.}
Figure~\ref{fig:russia-censored-destination} shows the respective probability for the destination ASes that are blocked by the Russian state.
As these websites have been predominantly blocked since the start of Russia's full-scale invasion of Ukraine,
we do not have any  data from 2020.
We refrained from measuring IPv6 as only two of the candidate ASes were IPv6-ready.
The ASes that are found towards these destinations are also found towards the Tranco List,
with a single exception: \textas{AS3223 VOXILITY}, an Internet infrastructure provider based in UK.

\begin{figure*}
	\centering
    \begin{subfigure}[b]{\textwidth}
        \includegraphics[width=\textwidth]{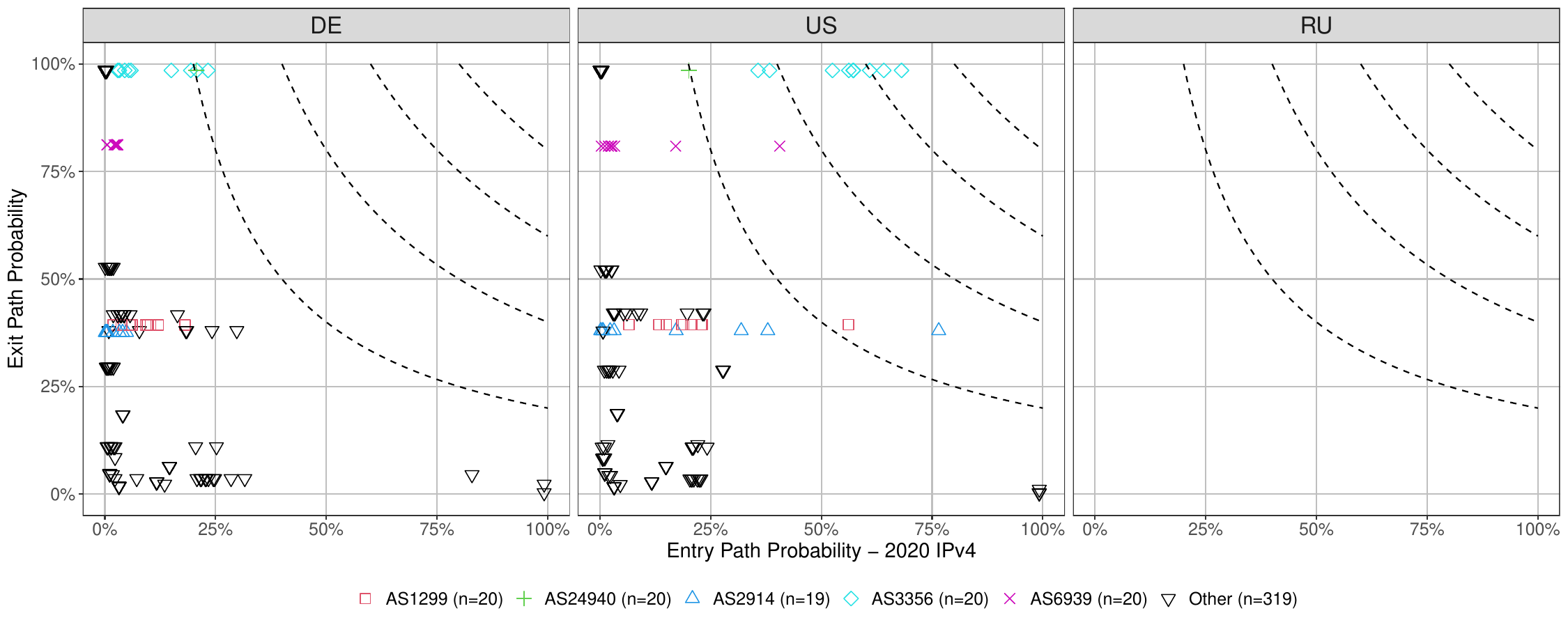}
        \caption{2020 IPv4}
        \label{fig:combined-2020-v4}
    \end{subfigure}

    \begin{subfigure}[b]{\textwidth}
        \includegraphics[width=\textwidth]{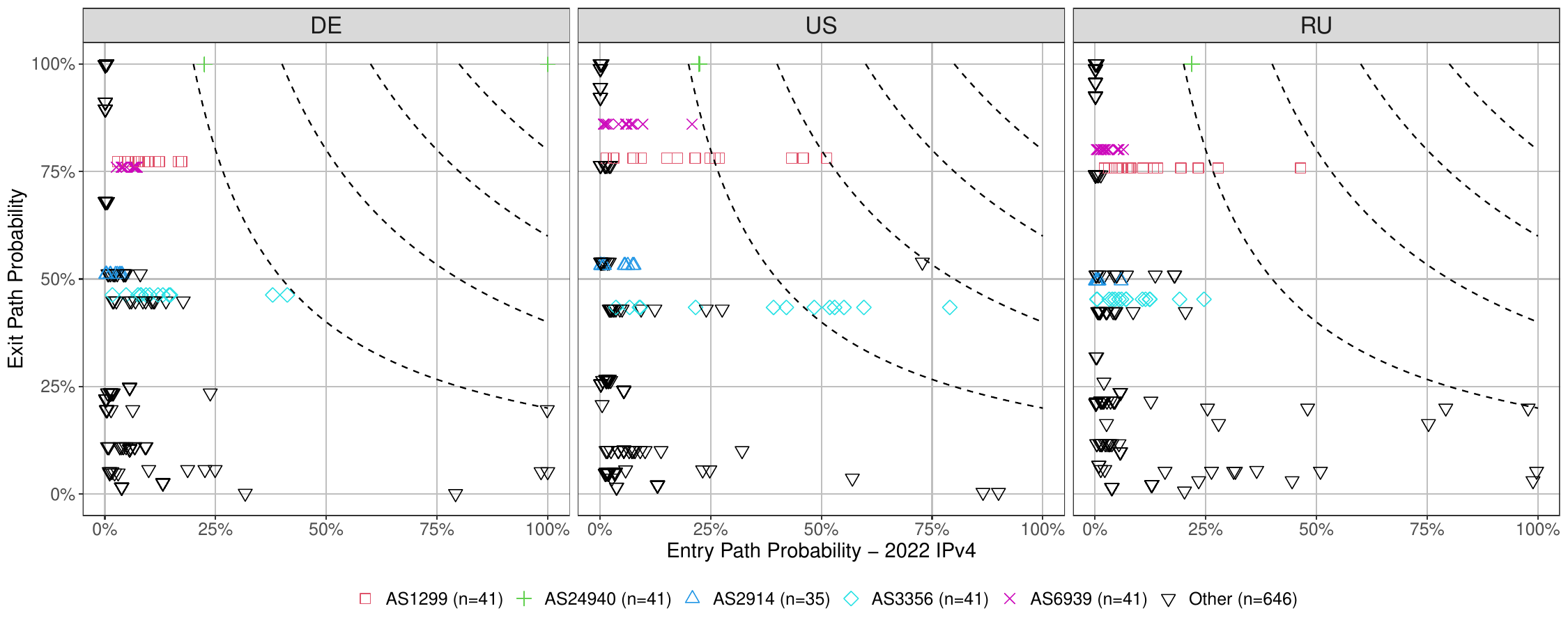}
        \caption{2022 IPv4}
        \label{fig:combined-2022-v4}
    \end{subfigure}

    \begin{subfigure}[b]{\textwidth}
        \includegraphics[width=\textwidth]{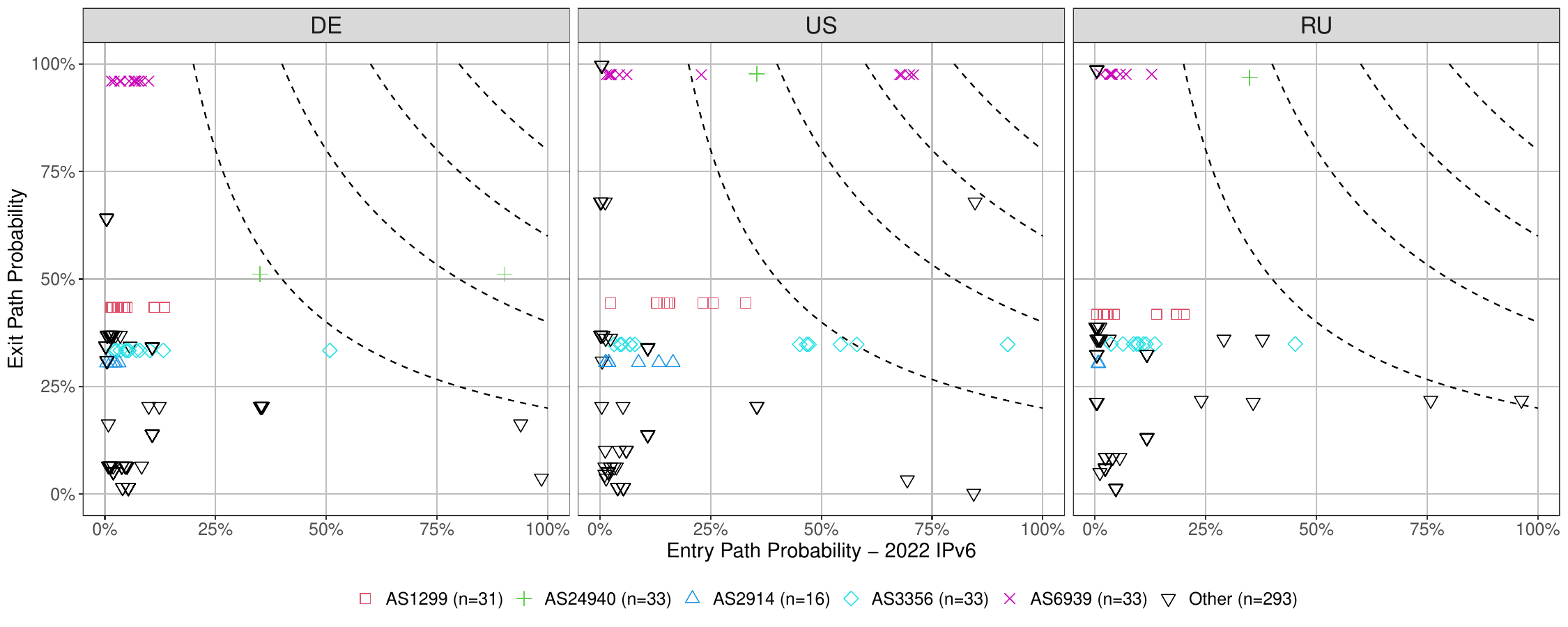}
        \caption{2022 IPv6}
        \label{fig:combined-2022-v6}
    \end{subfigure}

    \caption{ASes and their potential for traffic correlation for different years, protocols and countries for Tranco List destinations. Each data point represents an AS that appears on the entry and the exit path, and thus has the potential to perform traffic correlation. \fix{Contour lines at 20\%, 40\%, 60\% and 80\% highlight data points with highest combined probability.}}
    \label{fig:3x3combination}
\end{figure*}

\begin{figure} 
\centering
\includegraphics[width=.95\textwidth]{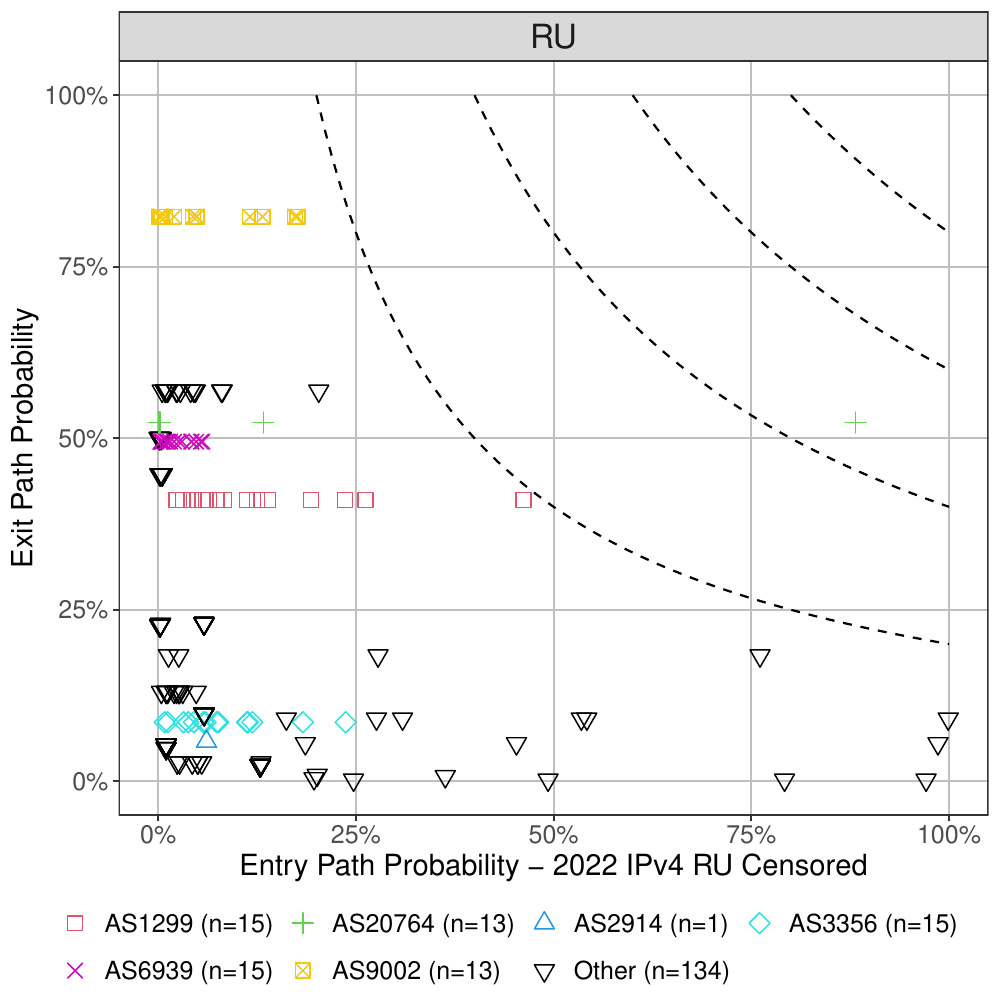}
\caption{ASes and their potential for traffic correlation for ASes hosting websites that are blocked by Russia. Each data point represents an AS that appears on the entry and the exit path, and thus has the potential to perform traffic correlation.}
\label{fig:result-combined-censored}
\end{figure}

\subsection{Potential ASes for Traffic Correlation}
\label{sec:evaluation_fourth}
As a final step, we combine the results from Tor's entry paths, between client and guard relays,
and exit paths, between exit relays and destinations.
We calculate the probability that an intermediate AS is residing on both paths as the latter is the prerequisite to conduct a successful correlation attack deanonymizing the client.

\textbf{Tranco List.}
Our results are depicted in Figure~\ref{fig:3x3combination},
providing the respective probability for the three measurements 2020 IPv4, 2022 IPv4, and 2022 IPv6,
as well as the three investigated countries Germany, the \typo{US}, and Russia.
Each data point in a graph represents a transit AS that has both entry and exit path probability higher than 0\%.
For the entry path, we show data points for all relevant clients.
For the exit path, we use the maximum probability of all measured destination ASes, which represents the worst-case scenario -- i.e., an attacker has the best \typo{chance} to correlate traffic when this target is visited by the Tor user.

In summary, \textas{AS24940 HETZNER} is strong in all scenarios:
First, it serves destinations and is thus likely to be on the exit side.
Second, it hosts a high share of guard bandwidth,
and is thus likely to serve as a guard relay,
eventually appearing on the entry side.
In combination, this leads to a high chance of being capable to correlate Tor traffic.
As an exception, the measurement on the bottom left (2022 IPv6 DE) shows a reduced exit probability for \textas{AS24940 HETZNER}.
In this case, the selected measurement probe for scheduling traceroutes from \textas{AS24940 HETZNER} to the Tor network -- corresponding to (3) in Figure~\ref{fig:scans} -- ran into a timeout and did not yield any results. %
Although we also measure the same routes in the opposite direction -- i.e., from the Tor network to \textas{AS24940 HETZNER}, corresponding to (2) in Figure~\ref{fig:scans} -- this only covered about 50\% of AS probability, due to the lack of RIPE Atlas availability in exit relay ASes (cf. Figure~\ref{fig:as-guard-propability}).

Since 2020, the exit probability of \textas{AS3356 LEVEL3} has decreased substantially. %
With this, its overall chance for a successful attack decreased,
both for German-based and US-based clients.
Yet, this AS has still been considered relevant due to having a good probability to be found on the entry side, particularly in the \typo{US}.
In return, \textas{AS1299 TWELVE99} has increased its exit probability in this time span.

For IPv6, we see high chances for US-based traffic to be correlated:
\textas{AS6939 HURRICANE} stands out.
It has high exit probability and also respectable entry probability at several client ASes.
Beyond, \textas{AS3356 LEVEL3} is noteworthy because it has an excellent entry probability for specific client ASes.
For both protocols,
it appears that there are less chances to correlate traffic originating from Russian-based client ASes.

\pagebreak
\textbf{Destinations Blocked by Russia.}
The combined probabilities for the ASes hosting websites that are blocked by the Russian state are presented in Figure~\ref{fig:result-combined-censored}.
\fix{Again, the contour lines highlight data points at 20\%, 40\%, 60\% and 80\% combined probability. }
In this case, we see again that there are lower chances to correlate Russian-originating traffic than those from other countries.
Although the client ASes (Russia) are within regional proximity to our destination ASes (Russia, Ukraine), the relevant transit ASes do not change much from our previous results.
As an outlier, \textas{AS20764 RASCOM} appears with \fix{an exit path} probability of 52.3\%.
\fix{It is } a consequence of the client AS simultaneously being the transit of the destination
and \typo{appearing} for a single \fix{client (\textas{AS20764} itself)} only.

Consequently, we assume that a regional attacker (e.g., a nation-state) is not able to match entry- and exit packets of local Tor clients.

\section{Discussion} \label{sec:discussion}
Adversaries residing along the path to/from a guard relay and from/to an exit relay bear the potential to correlate traffic, thus defeating the very goal of the anonymization network Tor. 
In this paper, we applied our previously developed measurement methodology~\cite{mayer2020actively} 
-- capable to detect such potentially malicious players --  to additional scenarios.
In particular,
we (a) repeated our measurements from 2020 to observe changes over time,
(b) adopted our approach for IPv6 to analyze the threat when using this next-generation Internet protocol,
and (c) extended our client- and destination sets to investigate the current situation in Russia
where censorship intensified after its full-scale invasion of Ukraine, starting on February 24th, 2022.

\textbf{Development over Time and Protocols.}
Our work does not provide any new impending AS-level adversaries.
The probability of an AS to be on the entry side and/or on the exit side is -- apart from a handful of changes -- stable over time (2020 and 2022) and protocol versions (IPv4 and IPv6).
This is good news:
The Tor network and also the underlying routing structure of the Internet remain to a large extent stable.
Tor is able to provide anonymity to users at a constantly high quality;
however, targeted attacks against hand-picked combinations of clients and destinations in close proximity cannot be fully prevented (e.g., \textas{RASCOM}).
Beyond, it means that active measurements like ours do not necessarily have to be performed on a daily basis
-- longer intervals are fine, reducing the effort for measurements.

\textbf{Division of Roles.}
Major transit ASes like \textas{HURRICANE} or \textas{LEVEL3} are the prime suspects
and are indeed capable of performing traffic correlation for many combinations of client and destination.
In addition, we identified networks simultaneously serving multiple roles, which puts them in a good position for correlation attacks.
For example, the data center operator \textas{HETZNER} serves as a hosting provider for many destinations (e.g., major websites);
at the same time, it hosts a high amount of guard relays.
In total, they account for 22.4\% of the guard bandwidth.
This puts the AS in a favorable position for correlation attacks:
The AS is likely to be part of a Tor path's entry side due to the many guard relays,
and there is a high chance of it being included in the exit path due to the many hosted destinations.
An operator of a \textas{HETZNER}-based guard relay even found that 15\% of the relay's traffic is forwarded to a relay within the same AS~\cite{torhetznerrelays2022}.

Ideally, guard relays should be -- in network terms -- close to the clients (e.g., in an ISP),
and the exit guards close to the destination (e.g., in a data center),
meaning that \textas{HETZNER} would be a good candidate to operate exit nodes.
We suggest to take this into account when deploying new guard- and/or exit relays,
either as a private individual or an organization.
An AS-aware circuit selection algorithm of Tor might also be beneficial
but bears the risk that the chosen ASes allow to trace it back to the origin,
see Section~\ref{sec:relatedwork} on Related Work.
Finally, we argue for increased AS diversity in the Tor network. 
Even with simple measurements, we see that the distribution of Tor relays is skewed. 
We hope that our measurements can improve an informed decision of how this diversity should be achieved.

\textbf{Russia.}
Since its full-scale invasion of Ukraine,
Russian state authorities are blocking access to online information that is not in line with the official reports.
This includes, among others, social networks, as well as local and independent media outlets.
Censorship might be overcome using Tor, and our measurements show that the chance of deanonymization due to traffic correlation is low for Russian users.
In fact, it is even lower than for users in Western democracies like Germany or the \typo{US} (in which information censored in Russia is accessible anyway).
Beyond, ASes that have the potential to perform successful correlation attacks are operated by companies in Western countries, further reducing the risk for Russian users.
At the moment, however, the main challenge is to access Tor: 
Russian authorities aim to block guard relays, thus hindering the technology's use.
The Tor community puts in a lot of effort to stay ahead of governmental blocking strategies~\cite{dingledine_russiacenscor_2022}.

\textbf{Open Source.}
We publish our source code openly available. 
This enables other entities such as large relay operators to also perform measurements.
All measurement results gathered through RIPE Atlas are openly available as well and could include valuable results for the Tor network.
We argue that large relay operators should deploy RIPE Atlas probes in their networks, not only to further improve our (future) results but also to enable other measurements.
Just a few more probes would increase the coverage significantly.

\fix{
\section{Limitations and Future Work}\label{sec:limitations}
}

\fix{
\textbf{AS Coverage.}
Our traceroute measurements are limited by the current AS-level coverage\footnote{\url{https://sg-pub.ripe.net/petros/population_coverage/table.html}} of the RIPE Atlas platform.
While RIPE Atlas provides considerable coverage of a country's Internet users for Western countries (e.g., 92\% in Germany and 86\% in the US), its scope in illiberal or censoring states is often constrained. 
For example, the coverage, at the time of our measurements, was 26\% in Russia,
declining from 60\% in 2020.
Due to the current geo-politic situations and lacking alternatives, 
we nevertheless opted to for the inclusion of Russia as our case study.
In comparison to Russia,
China's Internet population is covered well by RIPE Atlas (83\%),
and renders it a candidate for further studies.
Additionally, revisiting our measurements with increased IPv6 coverage and support among Tor relays could yield interesting results in the future.
}

\fix{
\textbf{Selection of Client and Destination ASes.}
Since tracerouting all possible client and destination ASes was not feasible,
we had to limit our measurements to a subset of ASes.
The chosen AS sets are intended to reflect the reality best possible, i.e., the client sets should match ASes that contain actual Tor users and the destination sets destinations that are actually requested via Tor.
A simple way to determine these ASes would be to capture traffic from (self-hosted) Tor relays;
this, however, raises ethic concers due to snooping on Tor users and we used popular  client and destination ASes instead.
For our case study of Russia's full scale invasion of Ukraine, 
we used destinations that are blocked by the Russian regulator Roskomnadzor.
We expect these destinations to be accessed via Tor as Russian Internet users cannot access them in a regular way;
thus, we believe this destination set to be closer to reality than the others.
Yet, there are no figures supporting this belief available.
}

\fix{
\textbf{Adversary Granularity.}
While this study specifically looks for adversaries at the granularity of ASes, 
there are other ways to group entities that could perform traffic correlation attacks.
In some cases, organizations act as multiple ASes which means that the results (i.e., probabilities) of these ASes from our measurements have to be cumulated.
Additionally, powerful nation states or intelligence agencies could force compliance of ASes within their jurisdiction to form an even more potent adversary.
Finally, we executed a single traceroute for each AS pair to determine traffic routes.
Future research could provide more precise results by doing this in a more fine-grained manner, e.g., by measuring routes from different network prefixes or regions for every selected AS.
}

\fix{
\textbf{Simplified Tor Model.}
Our study is based on the traditional model of Tor covering only publicly known guard- and exit relays.
In practice, Tor's architecture is constantly updated to cope with the ongoing censorship efforts of nation states like China or Russia.
Therefore, Tor  has introduced modular ``plugable transports'' (e.g., obfs4 bridges, Snowflake proxies) serving as relays which are not publicly known.
This makes it harder to block these relays.
We speculate that these add-ons could have positive effects on the AS distribution of the entry nodes (cf. \emph{Division of Roles} in Section~\ref{sec:discussion}) due to being more lightweight, ephemeral, and easy to set up by inexperienced users (e.g., via a browser plugin).
We consider this aspects to be part of future research.
}

\section{Conclusion}\label{sec:conclusion}
We applied our measurement technology,
which was developed in preliminary work~\cite{mayer2020actively},
to additional three use cases.
This line of action allowed us to get a broader picture of current deanonymization attacks in the Tor network,
and to infer actors with the potential to do so.
In particular,
we (a) repeated our measurements from 2020 to observe changes over time for users in Germany and the \typo{US},
(b) adopted our approach for IPv6 to analyze the threat when using this next-generation Internet protocol,
and (c) investigated the current situation in Russia where censorship has been intensified \fix{with} the beginning of its full-scale invasion of Ukraine on February 24th, 2022.

We indeed identified a small set of ASes with the potential to perform deanonymization attacks.
Most of them are large transit providers,
but we also found an AS which simultaneously hosts high numbers of destinations and Tor guard relays.
Hence, this AS has a high chance to appear on a Tor circuit's entry- and exit path, and consequently, successfully conducting traffic correlation to deanonymize individual Tor users.
\fix{Once again, this exposes the problems of centralization and shows that there is room for improvement regarding the placement of guard-, and exit relays on the Internet.}
The former should be close to the clients, the latter close to the destinations.

While the numbers of individual ASes have changed since 2020,
the overall picture does not reveal a significant change for Tor users in Germany and the \typo{US}.
Just as little does the protocol choice, i.e., IPv4 or IPv6,
have a significant impact.
We conclude that the Tor network provides anonymization to its users at a consistent quality.
According to our results, 
Russian users are even less prone than Western ones to become deanonymized.
Tor allows the former to securely access popular international websites as well as websites that have been censored.
Beyond, the few ASes with the potential to perform successful deanonymization attacks are operated by Western companies,
further reducing the risk for users in Russia.

\section*{Acknowledgments}
We want to thank David Schmidt for his preliminary work on this topic.
This material is based upon work partially supported
by (1) the Christian-Doppler-Laboratory for Security and
Quality Improvement in the Production System Lifecycle;
the financial support by the Austrian Federal Ministry for
Digital and Economic Affairs, the National Foundation for
Research, Technology and Development and the Christian
Doppler Research Association are gratefully acknowledged;
(2) SBA Research (SBA-K1), a COMET Centre within the
framework of COMET – Competence Centers for Excellent Technologies Programme and funded by BMK, BMDW,
and the province of Vienna. The COMET Programme is
managed by FFG; (3) Project 877110 2big2fail funded by
the Program "BRIDGE 1" (FFG); (4) Project DynAISEC
FO999887504 funded by the Program "ICT of the Future"
– an initiative of the Austrian Ministry of Climate Action,
Environment, Energy, Mobility, Innovation and Technology.

\printcredits

\bibliographystyle{cas-model2-names}

\bibliography{ripetor.bib}

\appendix
\section{Client and Destination AS Sets}
\label{sec:appendix-client-and-destination-sets}

\subsection{Client Sets}
\noindent
$C_{2020-DE-v4}$ = {\small \{AS3320, AS6830, AS31334, AS8881, AS3209, AS6805, AS553, AS680, AS8422, AS9145\}}
\\
$C_{2020-US-v4}$ = {\small \{AS7922, AS701, AS7018, AS209, AS20115, AS22773, AS5650, AS20001, AS10796, AS11427\}}
\\\\
$C_{2022-DE-v4}$ = {\small \{AS3320, AS3209, AS8881, AS6805, AS553, AS680, AS60294, AS24940, AS8422, AS9145\}}
\\
$C_{2022-US-v4}$ = {\small \{AS7922, AS7018, AS701, AS209, AS20115, AS22773, AS5650, AS20001, AS47583, AS20473\}}
\\
$C_{2022-RU-v4}$ = {\small \{AS12389, AS8402, AS25513, AS42610, AS35807, AS12714, AS3216, AS8359, AS12668, AS31200\}}
\\\\
$C_{2022-DE-v6}$ = {\small \{AS3320, AS3209, AS8881, AS6805, AS8422, AS199284, AS60294, AS24940, AS8767, AS680\}}
\\
$C_{2022-US-v6}$ = {\small \{AS7922, AS7018, AS701, AS47583, AS20473, AS62538, AS20001, AS209, AS22773, AS20115\}}
\\
$C_{2022-RU-v6}$ =  {\small \{AS42610, AS25513, AS202422, AS8331, AS12668, AS20764, AS50716, AS35807, AS12714, AS15974\}}

\subsection{Destination Sets}
\noindent
$D_{2020-v4}$ = {\small \{AS3, AS15169, AS4837, AS24940, AS36351, AS14618, AS16509, AS14907, AS3356, AS7941\}}
\\\\
$D_{2022-TRANCO-v4}$ = {\small \{AS15169, AS16509, AS8075, AS4837, AS14907, AS55990, AS37963, AS132203, AS4134, AS4812, AS47764, AS29169, AS14618, AS396982\}}
\\
$D_{2022-TRANCO-v6}$ = {\small \{AS15169, AS16509, AS14907, AS47764, AS63949, AS3, AS37963, AS197695, AS32, AS14618\}}
\\\\
$D_{2022-RU-CENSORED-v4}$ = {\small \{AS200350, AS15497, AS25532, AS207651, AS9123, AS28907, AS3326, AS197695, AS25521, AS12722\}}

\clearpage

\bio{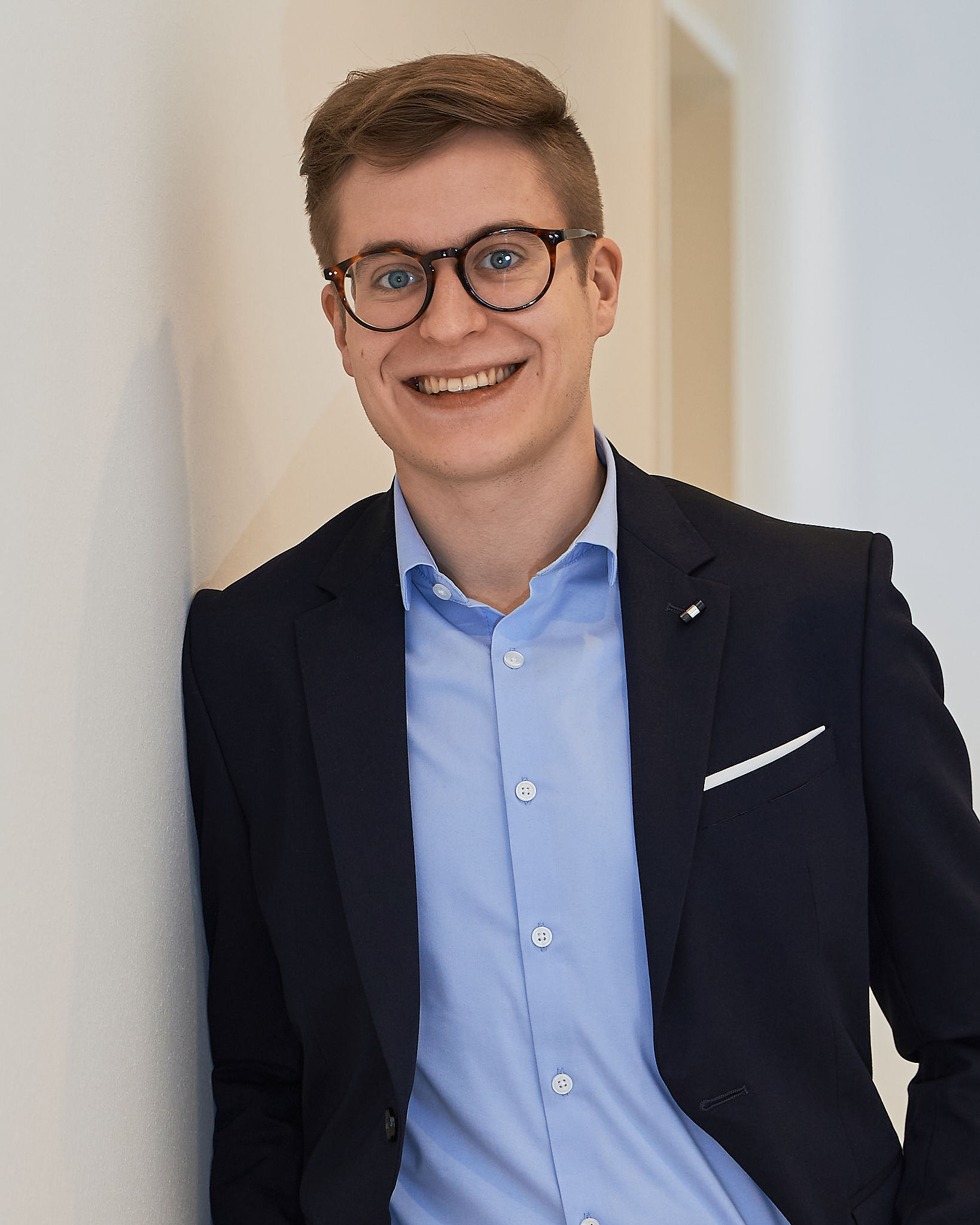}
\textbf{Gabriel K. Gegenhuber} is a PhD student at University of Vienna, Austria, Research Group Security and Privacy.
Gabriel received a BSc in Software \& Information Engineering and an MSc in Software Engineering \& Internet Computing at the Technical University of Vienna.
His research interests include mobile networks, network measurements, network security, and privacy-enhancing technologies.
\endbio

\bio{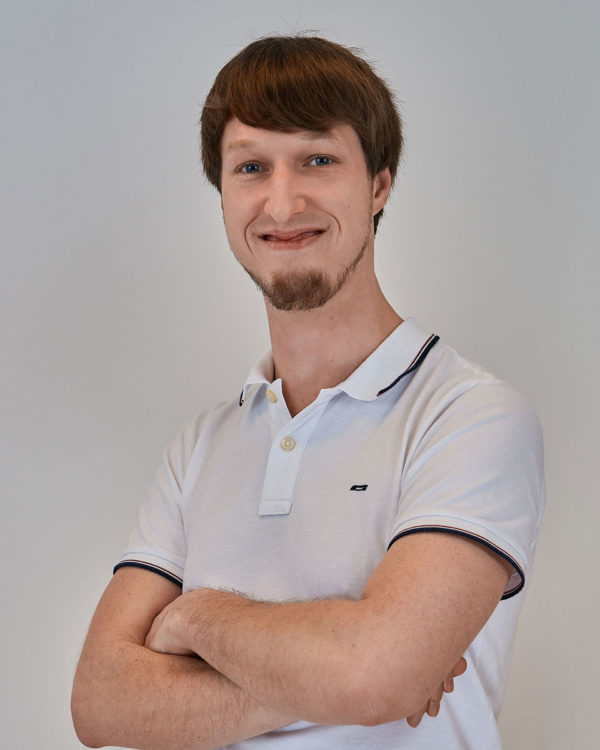}
\textbf{Markus Maier} is a PhD student at University of Vienna, Austria, Research Group Security and Privacy.
He received his BSc and MSc at Technical University of Vienna in Software Engineering \& Internet Computing. 
His research interests include routing, network measurement and network security.
\vspace{0.7cm}
\endbio

\bio{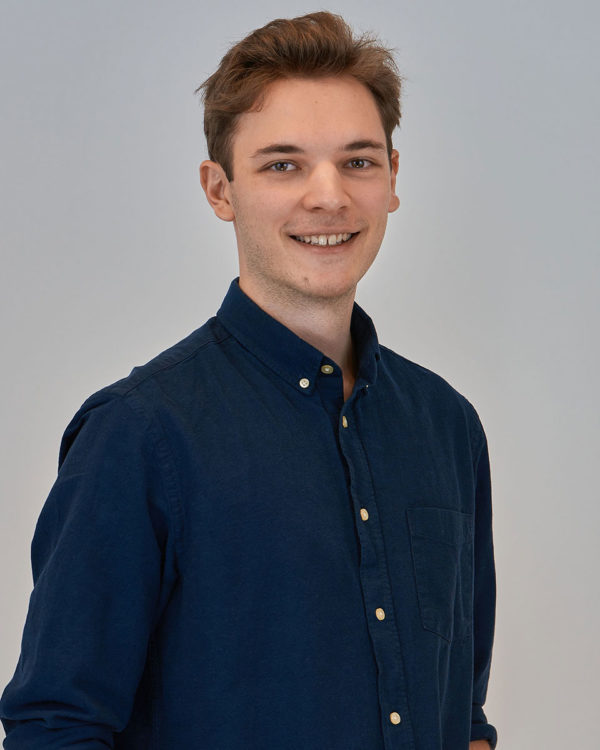}
\textbf{Florian Holzbauer} is a PhD student at University of Vienna, Austria, Research Group Security and Privacy. He received his BSc and MSc at University of Applied Sciences St.Pölten. His research focuses on Internet measurements. He detected flaws in email-related protocol adoption and is currently looking for flaws in IPv6 deployments. 
\vspace{0.7cm}
\endbio

\bio{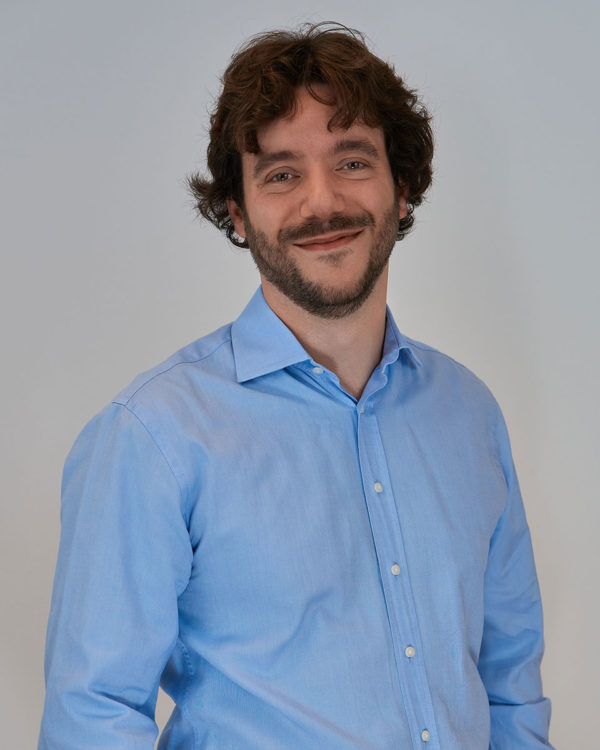}
\textbf{Wilfried Mayer} received a master’s degree in Software Engineering and Internet Computing, and a doctoral degree in computer science at TU Wien. His research interests are focused on measuring privacy-enhancing technologies.
\vspace{1.5cm}
\endbio

\bio{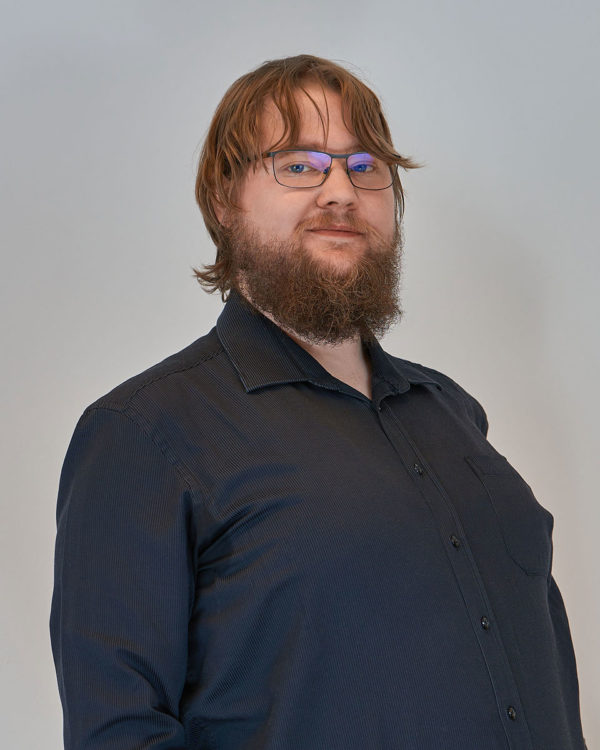}
\textbf{Georg Merzdovnik} received a BSc in computer engineering, an MSc in software and information engineering, and a PhD in computer science with distinction at TU Wien. Currently, he leads the research group on Systems and (I)Iot Security at SBA Research. 
Georg’s research interests include applied systems and software security, IoT security (ranging from device to network level) as well as online privacy in general.
\endbio

\bio{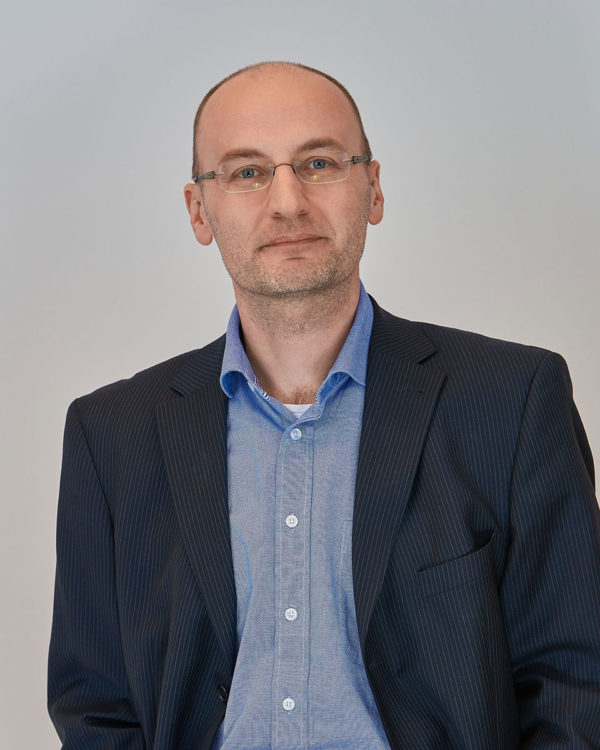}
\textbf{Edgar Weippl} Edgar graduated with a PhD from TU Wien. Afterwards, he was an assistant professor at Beloit College, WI, and a consultant for the software vendor ISIS Papyrus in New York, NY, Albany, NY, and Frankfurt, Germany. Returning to Vienna, he co-founded the research center SBA Research in 2004. In 2020, Edgar accepted a position as full professor at the University of Vienna.
\vspace{1.2cm}
\endbio

\bio{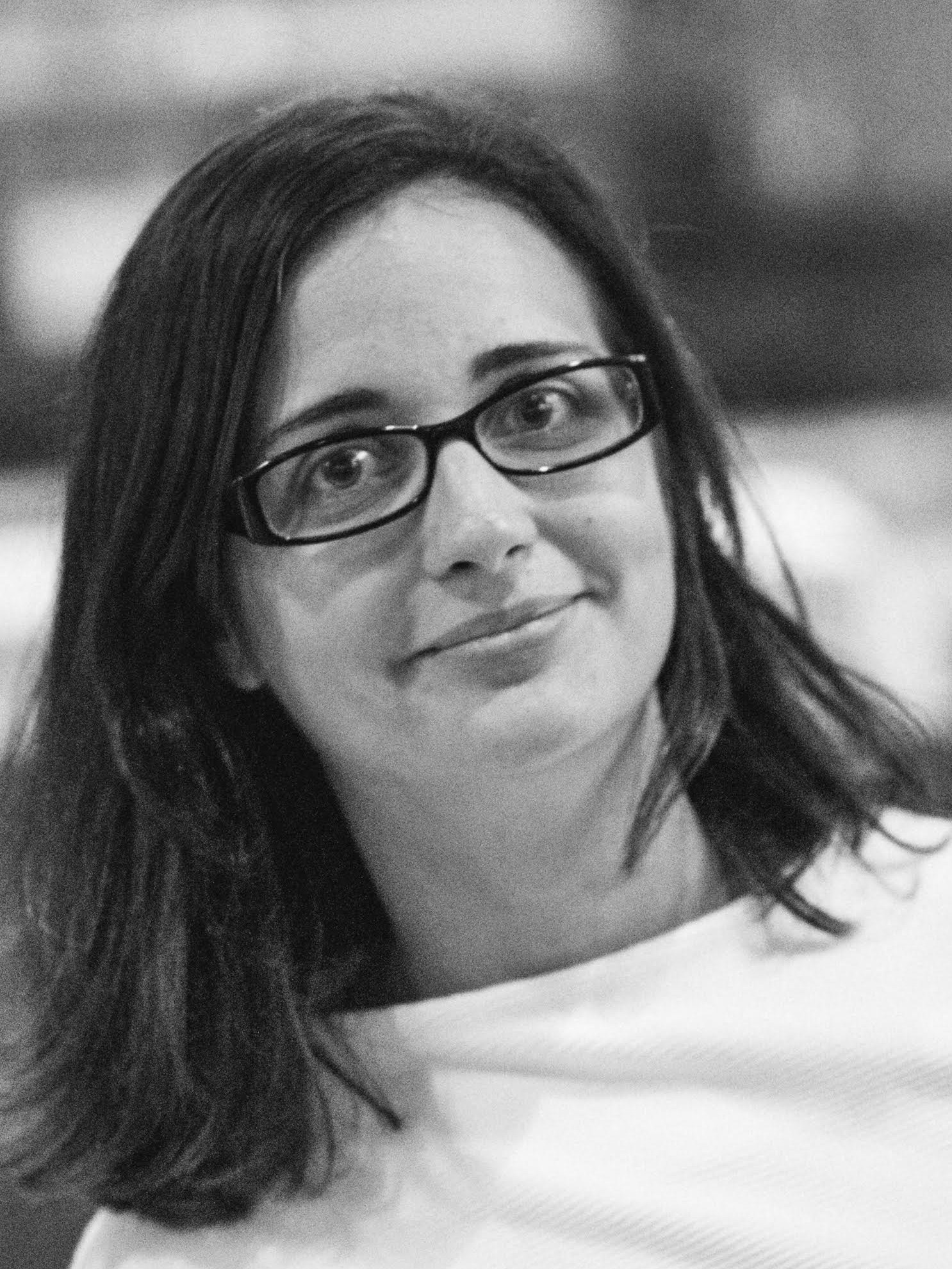}
\textbf{Johanna Ullrich} received a Ph.D. sub auspiciis praesidentis from TU Wien.
Currently, she is a key researcher at SBA Research, Austria, leading the Networks and Critical Infrastructures Security Research Group, and a researcher of the Christian Doppler laboratory SQI.
She was awarded the Research Prize of the Dr. Maria Schaumayer Foundation and nominated for the Hedy Lamarr Prize twice. 
Her research focuses on network security, particularly measuring experiments and IPv6. 
\endbio

\end{document}